\documentclass[aps,prb,twocolumn,showpacs,superscriptaddress]{revtex4-1}
\usepackage[colorlinks,linkcolor=blue,anchorcolor=blue,citecolor=blue,urlcolor=blue]{hyperref}
\usepackage{graphicx}
\usepackage{epstopdf}
\def\Im{\rm{Im}}
\def\be{\begin{equation}} \def\ee{\end{equation}}
\def\bea{\begin{eqnarray}} \def\eea{\end{eqnarray}}

\def\nn{\nonumber}

\begin{document}
\title{Finite-temperature valence-bond-solid transitions and
thermodynamic properties of interacting SU($2N$) Dirac fermions}
\author{Zhichao Zhou}
\affiliation{School of Physics and Technology, Wuhan University, Wuhan
430072, China}
\author{Da Wang}
\affiliation{National Laboratory of Solid State Microstructures and
 School of Physics, Nanjing University, Nanjing, 210093, China}
\author{Congjun Wu}
\affiliation{Department of Physics, University of California, San Diego,
CA 92093, USA}
\author{Yu Wang}
\email{yu.wang@whu.edu.cn}
\affiliation{School of Physics and Technology, Wuhan University, Wuhan
430072, China}

\begin{abstract}
We investigate the SU($2N$) symmetry effects with $2N>2$ on the
two-dimensional interacting Dirac fermions at finite temperatures,
including the valence-bond-solid transition, the Pomeranchuk effect,
the compressibility and the uniform spin susceptibility, by performing
the determinant quantum Monte Carlo simulations of the half-filled SU($2N$)
Hubbard model on a honeycomb lattice.
The columnar valence-bond-solid (cVBS) phase only breaks the three-fold discrete
symmetry, and thus can survive at finite temperatures.
The disordered phase in the weak coupling regime is the thermal Dirac
semi-metal state, while in the strong coupling regime it is largely a
Mott state in which the cVBS order is thermally melted.
The calculated entropy-temperature relations for various values of the
Hubbard interaction $U$ show that, the Pomeranchuk effect occurs when the specific entropy is below a characteristic value of $S^*$ --- the maximal entropy per particle from the spin channel of local moments.
The SU($2N$) symmetry enhances the Pomeranchuk effect, which facilitates the
interaction-induced adiabatic cooling.
Our work sheds new light on future explorations of novel states of
matter with ultra-cold large-spin alkaline fermions.
\end{abstract}
\maketitle

\section{Introduction}
The low-energy quasi-particles on a honeycomb lattice exhibit
the two-dimensional (2D) massless Dirac-fermion-type band structure.
The interplay between charge and spin degrees of freedom together with
the Dirac band structure brings novel features of quantum phases, which has
become a major research focus in condensed matter physics
since the discovery of graphene \cite{CastroNeto2009}.
The strong interaction effects in 2D Dirac fermion systems have been
investigated extensively by applying quantum Monte Carlo (QMC)
simulations to the SU(2) Hubbard model, a paradigmatic model for
Mott physics of interacting electrons \cite{Paiva2005,Meng2010}.
Because of the bipartite nature of the honeycomb lattice, it exhibits
the antiferromagnetic (AF) long-range order in the Mott-insulating phase.
The transition from the Dirac semi-metal phase to the AF insulating
phase is found to be continuous \cite{Assaad2013,Parisen2015,Otsuka2016}.

Dirac fermions are not unique to high energy and solid state systems,
and also can be realized in optical lattices loaded with ultra-cold
fermionic atoms.
Unlike spin-$\frac{1}{2}$ electrons in solids which usually possess
the SU(2) symmetry, ultra-cold fermions often carry large hyperfine spins.
As proposed early by one of the authors, Hu and Zhang \cite{Wu2003},
ultra-cold alkali and alkaline-earth fermions provide a new
opportunity to study high symmetries that are typically studied in
the high energy context.
For example, the simplest large-spin fermions of spin-$\frac{3}{2}$ in
optical lattices generically possess the high symmetry of Sp(4),
or, isomorphically SO(5), without fine-tuning \cite{Wu2003}.
If the interaction is spin-independent, the symmetry is enlarged to SU(4).
These high symmetries are expected to give rise to exotic quantum
phases difficult to access in solids, which provide important guidance
in analyzing novel many-body physics with multi-component
fermions \cite{Wu2006a}.
On the other hand, it has been pointed out that the alkaline-earth fermion
systems respect the SU($2N$) symmetry, owing to their closed shell
electron structure.
Their hyperfine spins are simply nuclear spins, and thus the inter-atomic
scatterings are spin-independent, leading to the SU($2N$)
symmetry \cite{Gorshkov2010,Wu2012}.
Excitingly, the recent rapid progress of ultra-cold atom experiments
have already realized these SU($2N$) symmetric systems
\cite{Taie2010,DeSalvo2010,Taie2012,Pagano2014, Hofrichter2016}.
The quantum degenerate temperatures have been reached in
alkaline-earth atoms with the large hyperfine spins, e.g., $^{173}$Yb
with SU(6) symmetry \cite{Taie2010} and $^{87}$Sr with the SU(10)
symmetry \cite{DeSalvo2010}.
Excitingly, an SU(6) Mott insulator has also been realized with
$^{173}$Yb atoms in the optical lattice and the Mott insulating
gap has been observed in the shaking lattice experiment \cite{Taie2012}.

It will be interesting to combine the SU($2N$) symmetry and the 2D
Dirac fermion together, which can be realized by loading large spin
alkaline-earth fermions into the honeycomb optical lattice, to
investigate novel physics absent in the SU(2) Hubbard model
of Dirac fermions.
In a recent paper \cite{Zhou2016}, we investigated the novel effects of
the SU($2N$) symmetries on quantum many-body physics of Dirac fermions,
including quantum magnetism and the Dirac semi-metal-to-Mott insulator
transitions, by performing the projector determinant quantum Monte
Carlo simulations of the half-filled SU($2N$) Hubbard model on a
honeycomb lattice.
We found that, fundamentally different from the usual SU(2)
Mott-insulating phase which exhibits the AF N\'{e}el ordering,
the SU(4) and SU(6) Mott-insulating phases are identified with the
valence-bond-solid (VBS) order.
Both the columnar VBS (cVBS) and the plaquette VBS (pVBS) break the
same type of symmetry and compete, and the ground states are found
to exhibit the cVBS order.
The nature of the Dirac semi-metal-to-cVBS order transition
has been analyzed at the mean-field level.
It unveils the possibility of an exotic 2nd order quantum phase
transition seemingly forbidden by the Ginzburg-Landau theory,
which is also investigated and confirmed by the renormalization
group analysis \cite{Roy2013, Li2015, Scherer2016, Jian2016}.
Besides, our mean-field analysis also points out that the semi-metal-to-cVBS
transition at finite temperatures is still the 1st order.

In this work, we investigate the thermodynamic properties of the 2D
SU($2N$) Hubbard model on the honeycomb lattice by employing the
unbiased non-perturbative determinant QMC simulations.
We focus on the thermal cVBS transition and the interaction-induced
Pomeranchuk effect of the SU($2N$) Dirac fermions.
Since the cVBS state only breaks the discrete symmetry of lattice
translation, it can survive at finite temperatures and the
transition to the disordered state can take place at finite temperatures.
The weak coupling Dirac-semi-metal regime and the strong coupling
disordered Mott-insulating state are connected at finite temperatures.
The finite-temperature simulation studies of Hubbard models
show that the Pomeranchuk effect occurs when the specific entropy is below a characteristic value of $S^*$,
where $S^*$ represents the maximum amount of
specific entropy carried by the spin channel.
The Pomeranchuk effect is dependent on the symmetry and the lattice
structure.
We have shown that the multi-component SU($2N$) Hubbard
model on a honeycomb lattice significantly facilitates the Pomeranchuk
effect, which starts with even relatively high entropies.
Other thermodynamic properties, including the onsite particle number
fluctuations, compressibility, and the uniform spin susceptibility,
are also analyzed.

The rest of the paper is organalized as follows.
In Sect. \ref{sect:model}, the model Hamiltonian and parameters for determinant QMC
(DQMC) are introduced.
In Sect. \ref{sect:PhaseTransFT}, the thermal VBS transition is studied.
The entropy and the on-site occupation number are studied in
Sect. \ref{sect:thermoProp}.
Subsequently in Sect. \ref{response}, the density compressibility
and uniform spin susceptibilities are investigated.
The conclusions are drawn in Sect. \ref{ConDiscu}.

\section{Model and Method}
\label{sect:model}

\subsection{The SU($2N$) Hubbard model}
At half-filling, the Hubbard model with SU($2N$) symmetry takes the
following form on a honeycomb lattice,
\bea
H&=&-t\sum_{i \in A, \hat e_j; \alpha}
\left(c_{i\alpha}^{\dagger}c_{i+\hat e_j, \alpha}+h.c.\right)
\nn \\
&+&\frac{U}{2}\sum_{i\in A \oplus B}\left(n_{i}-N\right)^{2},
\label{eq:hubbard}
\eea
where $A$ and $B$ denote two sublattices of the honeycomb lattice;
$\hat{e}_{j}$'s with $j=1,2,3$ are vectors connecting each site with its
three nearest neighbors; the spin index $\alpha$ runs from $1$ to $2N$
and $n_{i}=\sum_{\alpha}c^{\dagger}_{i\alpha}c_{i\alpha}$ is the particle number
operator on site $i$;
$t$ and $U$ are the nearest-neighbor hopping integral and the on-site
Coulomb repulsion, respectively.
The chemical potential $\mu$ vanishes in the grand canonical Hamiltonian
due to half-filling.
$U$ is defined in the following convention:
in the atomic limit, $t/U\rightarrow 0$, if a single fermion is
removed from one site and put on any other site on the half-filled
2D system, the energy cost of this charge excitation is $U$,
which is independent of $N$.

\subsection{The numerical method}
We employ the DQMC method based on Blankenbecler-Scalapino-Sugar
algorithm \cite{BSS1981}.
QMC is a widely used non-perturbative and unbiased numerical method
for studying 2D strongly correlated systems.
Compared to other methods, the major advantage is that it is scalable
to large sizes and capable of yielding asymptotically exact results
provided that the sign problem is absent.
The half-filled SU($2N$) Hubbard model is free of the sign-problem in
the bipartite lattices.
In the DQMC simulations of the SU($4$) and SU($6$) Hubbard models
with repulsive interactions, an exact Hubbard-Stratonovich (HS)
decomposition is performed in the density channel involving complex
numbers, which maintains the SU($2N$) symmetry during the HS
decomposition \cite{Wang2014}.
The method of the HS decomposition is explained in Appendix \ref{app:com}.

The parameters of the QMC simulations are presented below.
Unless specifically stated, the time discretization parameter $\Delta\tau$
is set to $1/30$ at least, ensuring the convergence of the second-order
Suzuki-Trotter decomposition.
The $2\times L\times L$ honeycomb lattice with $L=9$ is simulated under
the periodic boundary condition which preserves the translational symmetry.
The finite-size effect on the entropy-temperature relations is analysed in Appendix \ref{app:U0_entropy}.
For a typical data point, we use 10 QMC bins each of which includes 2000
warm-up steps and 8000 measurements.
To investigate the thermal phase transition by the finite-size scaling,
the $2\times L\times L$ honeycomb lattices
with $L=6,9,12,15$ are simulated with at least 20 QMC bins, each bin
containing 500 warm-up steps and 500 measurements.
In our simulations, the Hubbard $U$ and temperature $T$ are given
in the unit of $t$.

\begin{figure}[htb]
\includegraphics[width=0.99\linewidth]{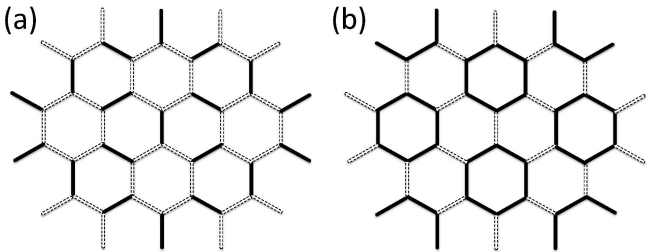}
\caption{The cVBS ($a$) and pVBS
($b$) configurations break the three-fold discrete symmetry,
and exhibit a $\sqrt{3}\times\sqrt{3}$ superunit cell. (This figure is taken from Ref. [\onlinecite{Zhou2016}].)
}
\label{fig:cpVBS}
\end{figure}

\begin{figure}[htb]
\includegraphics[width=0.99\linewidth]{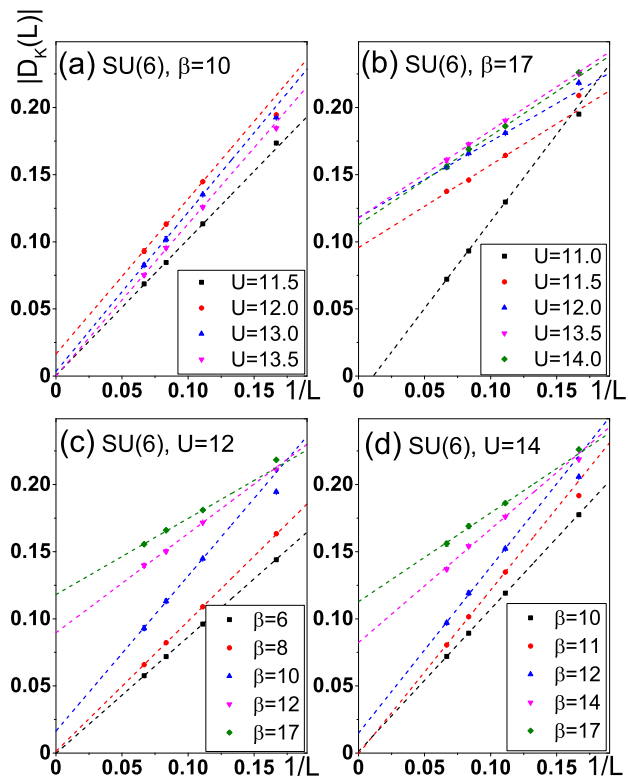}
\caption{Finite-size scalings of the VBS dimer parameter
$|D_{K}|$ for the half-filled SU($6$) Hubbard model at different
values of $U$ and $\beta$ close to the phase boundary:
($a$) $\beta=10$ with different values of $U$;
($b$) $\beta=17$ with different values of $U$;
($c$) $U=12$ with different $\beta$;
($d$) $U=14$ with different $\beta$.
The linear fitting is used starting from $L=9$, and error bars are
smaller than data points.
}
\label{fig:DK1}
\end{figure}

\subsection{The order parameters}
We define three bonds attached to site $i$ as
\bea
d_{i,\hat{e}_{j}} = \frac{1}{2N}\sum_{\alpha=1}^{2N}
(c^{\dagger}_{i,\alpha}c_{i+\hat{e}_{j},\alpha}+h.c.),
\eea
where $j=1,2,3$ represent three different bond orientations.
The cVBS and pVBS orders are defined in the same form as
\bea
D_{K}(L) = \frac{1}{L^2}\sum_{i\in A}
(d_{i,\hat{e}_{a}}+\omega d_{i,\hat{e}_{b}}+\omega^2 d_{i,\hat{e}_{c}})
e^{i\vec{K}\cdot \vec{r}_{i}},
\eea
where $\omega=e^{i\frac{2}{3}\pi}$ and $\vec{K}=(\frac{4\pi}{3\sqrt{3}a_{0}},0)$.
Their configurations are depicted in Fig. \ref{fig:cpVBS}.
Following Ref. [\onlinecite{Zhou2016}],
the difference between cVBS and pVBS can be distinguished through the
following parameter,
\bea
W=\int dz dz^* P(z,z^*) \cos 3\theta,
\eea
where $z=D_{K}$, $\theta=\arg(z)$, and $P(z,z^*)$ is
the density of probability that appears in  Monte Carlo samplings.
For the ideal cVBS and pVBS states without fluctuations,
$W$ equals $1$ and $-1$, respectively.
Certainly, fluctuations weaken the magnitudes of $W$, nevertheless,
the sign of $W$ can be used to distinguish whether the ordering
is cVBS or pVBS.
For the isotropic state, $W=0$.

\section{The finite temperature VBS transition}
\label{sect:PhaseTransFT}

\begin{figure}[htb]
\includegraphics[width=0.9\linewidth]{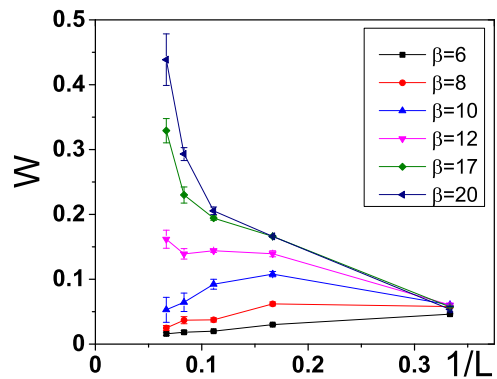}
\caption{Finite-size scalings of $W(L)$ for the half-filled
SU($6$) Hubbard model at $U/t=12$ with different values of
the inverse temperature $\beta$.
}
\label{fig:W}
\end{figure}

In Ref. [\onlinecite{Zhou2016}], the Dirac semi-metal-to-cVBS
transitions are shown to occur in the ground states of SU($4$) and SU($6$) Hubbard
models on a honeycomb lattice.
Quantum spin fluctuations are enhanced with increasing $2N$, and thus the strength of
cVBS order in the SU($6$) case is stronger
than that in the SU($4$) case.
In the SU($6$) case, the cVBS order starts to appear around
$U/t\approx 11$ and grows with increasing $U$ until reaching the peak value
around $U/t\approx 14$, and then decreases as $U$ further increases.
The cVBS state breaks the three-fold discrete symmetry exhibiting
the $\sqrt 3\times \sqrt 3$ structure, and thus the cVBS transition
should survive at finite temperatures.
In this section, we will further investigate the finite-temperature
VBS transitions of the SU($6$) Hubbard model.

The finite-size scalings of the VBS order parameters $|D_{K}|$ at fixed
temperatures with $\beta=10$ and $17$ are presented in Fig. \ref{fig:DK1}
($a$) and ($b$), respectively, where $\beta=t/T$.
At $\beta=10$, the system first undergoes a transition from the
disordered phase to the VBS phase and reenters the disordered phase, as Hubbard $U$ increases.
As shown in the analysis below, based on the calculation of $W$, the
nature of the thermal VBS phase is the same as that in the ground state --
the cVBS \cite{Zhou2016}.
The first transition located at $U/t\approx 11.5$ is a finite-temperature
version of the ground state Dirac semi-metal-to-cVBS transition, while the
second transition located around $U/t\approx 13.5$ is the thermal
melting of the cVBS state in the Mott-insulating background.
At a lower temperature with $\beta=17$, the VBS order strengths are still
non-monotonic with $U$, and this feature persists into the
ground state as shown in the previous zero-temperature projector DQMC
simulations \cite{Zhou2016}.
The reason is that, the reduction of the bond kinetic energy scale with
increasing $U$ suppresses the strength of VBS ordering.
In Fig. \ref{fig:DK1} ($c$) and ($d$), the finite-size scalings of cVBS
order parameters $|D_{K}|$ are shown for fixed values of $U$.
As temperature decreases, the VBS order develops.
At $U/t=12$, the critical temperature $T_c$ of the cVBS transition
is located in the range of $8<\beta_c<10$, or, $1/8>T_c/t>1/10$.
Similarly, at $U/t=14$, $11<\beta_c<12$.

In order to determine the type of the VBS state, we present the
finite-size scaling of $W$ for $U/t=12$ in Fig. \ref{fig:W}.
At $\beta<10$, $W$ approaches zero in the thermodynamic limit, which
signifies an isotropic disordered phase.
At $\beta>12$, $W$ has already developed a positive value in the
thermodynamic limit, which indicates a cVBS ordered phase.
The cVBS transition temperature $T_c$ based on the scaling of $W$
is in agreement with that based on the scaling of the VBS dimer parameter.

\begin{figure}[htb]
\includegraphics[width=0.9\linewidth]{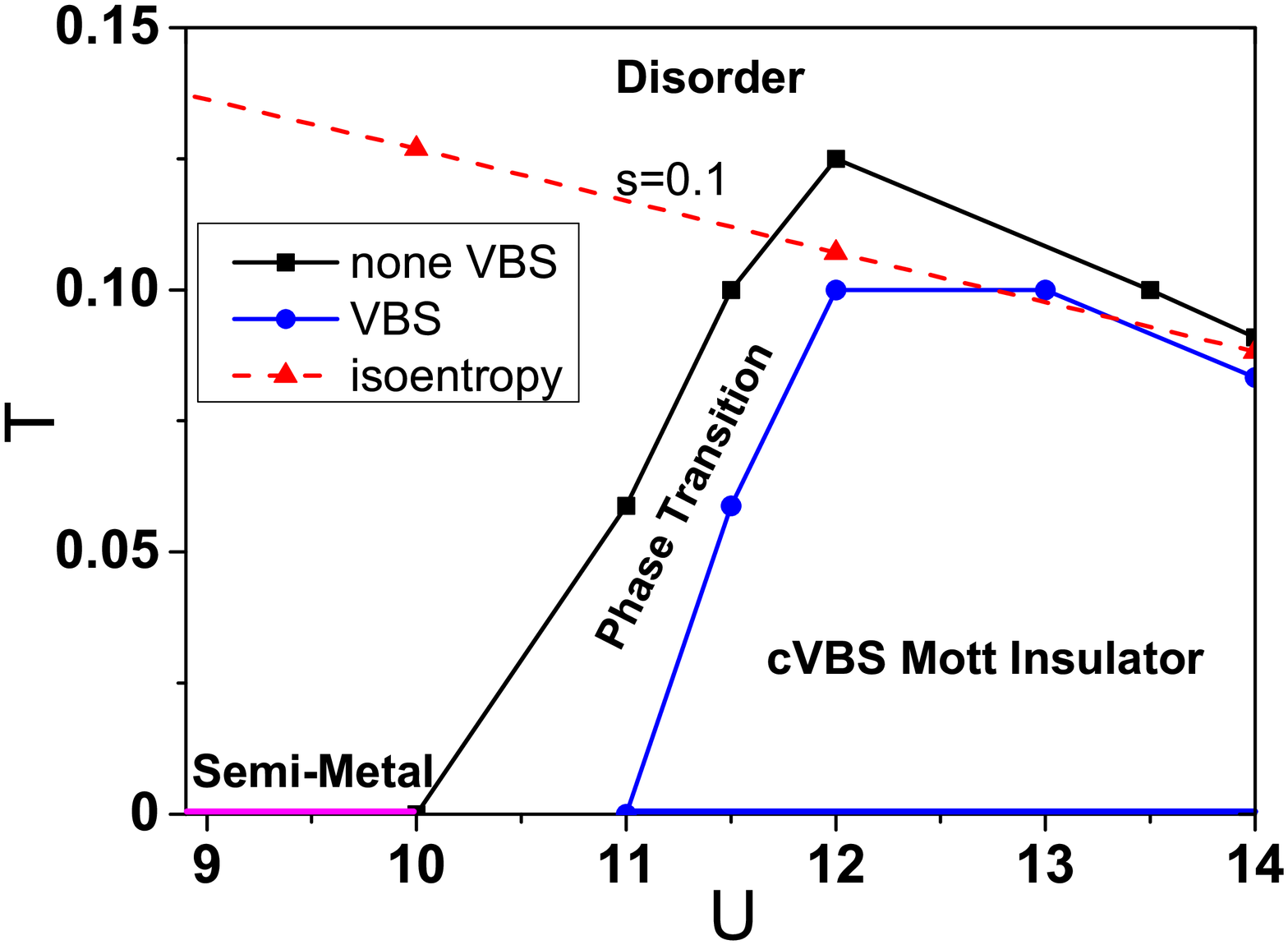}
\caption{The finite-temperature phase diagram of the half-filled
SU(6) Hubbard model on a honeycomb lattice.
The black and the blue lines represent the upper and lower
boundaries of the transition temperatures determined by
the DQMC simulations.
The zero-temperature results are extracted from Ref. [\onlinecite{Zhou2016}].
With denser $U$ and $T$, the two boundaries should merge into one.
The red dashed line represents the isoentropy curve
of $S_{SU(6)}=0.1$.
}
\label{fig:phase}
\end{figure}

Based on the above analysis, the finite-temperature phase diagram
of the SU($6$) honeycomb-lattice Hubbard model is plotted
in Fig. \ref{fig:phase}.
The transition temperature $T_{c}$ increases with $U$ in the interaction
range $11<U/t<12$, while it decreases as $U$ further increases.
The non-monotonic dependence of $T_c$ on $U$ is consistent with
the behavior of the cVBS ordering strength in the ground state, which
first increases until reaching the maximum, and then decreases as $U$ further increases.
The phase diagram Fig.\ref{fig:phase} suggests that, an SU($6$) symmetric
Mott insulator with the long-range cVBS order can be formed with an
atomic Fermi gas of $^{173}$Yb with the hyperfine spin $I=5/2$, if the
ultra-cold fermions on an optical honeycomb lattice are cooled down to
temperature regime below $T/t=0.1$. Simulations performed in the next section shows that
this temperature can be achieved by adiabatically increasing the interaction $U$ along the isoentropy curve of $s/k_B\approx 0.1$.

\section{The Pomeranchuk effect}
\label{sect:thermoProp}
In this section, we demonstrate, by means of the DQMC simulation, the
pronounced Pomeranchuk effect in the half-filled SU($4$) and SU($6$)
Hubbard models on a honeycomb lattice.

\subsection{The entropy-temperature relations}
\label{entropy}

\begin{figure}[htb]
\includegraphics[width=0.9\linewidth]{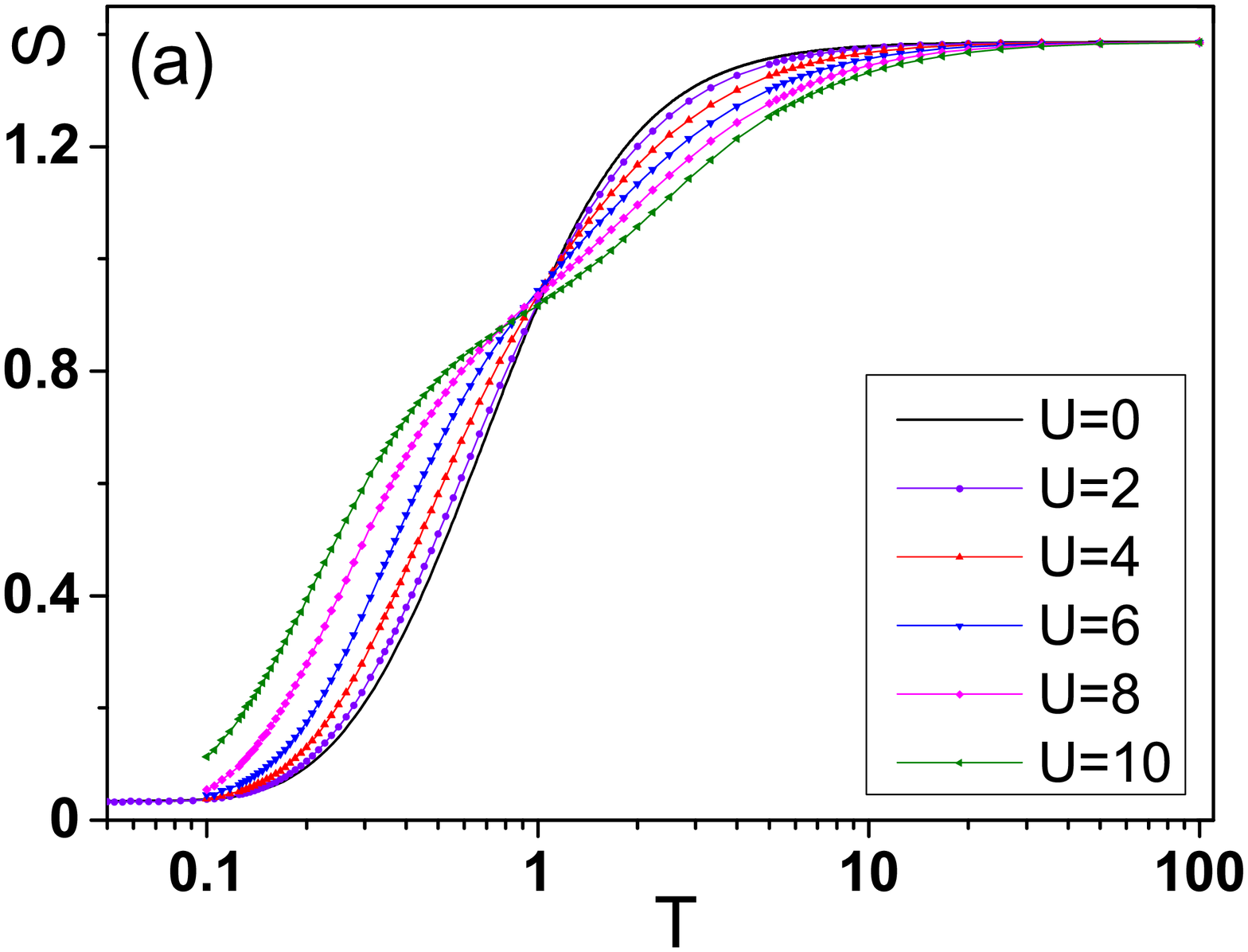}
\includegraphics[width=0.9\linewidth]{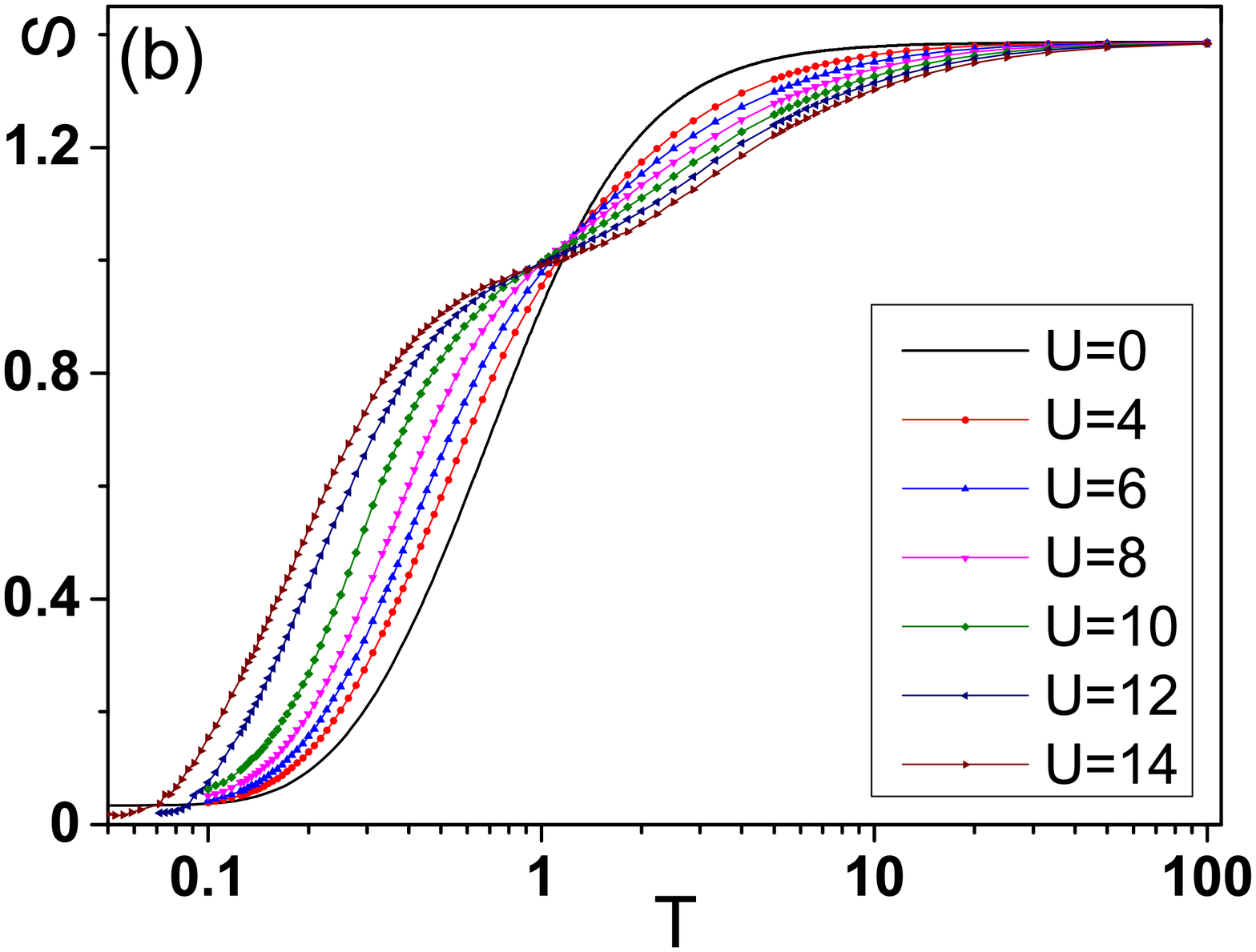}
\caption{The entropy per particle as a function of $T$
 at different values of $U$ in ($a$) SU($4$) and ($b$) SU($6$)
Hubbard models. Note that the $S-T$ curve for $U=0$ is calculated in Appendix \ref{app:U0_entropy}. The lattice size is $L=9$.
}
\label{fig:entropy}
\end{figure}

In ultra-cold atom experiments, entropy, rather than temperature, is
a directly measurable physical quantity \cite{Bloch2008}.
We present below the entropy-temperature relations in the half-filled
SU($4$) and SU($6$) Hubbard models on a honeycomb lattice.
The entropy per particle can be calculated by:
\bea
\frac{S(T)}{k_{B}}=\frac{S(\infty)}{k_{B}}+\frac{E(T)}{T}
- \int^{\infty}_{T}dT' \frac{E(T')}{T'^{2}},
\eea
where $E(T)$ is the internal energy per particle at temperature $T$.
In the high temperature limit, there are $2^{2N}$ possible states on each
site, and thus
$S(\infty)=k_B\frac{\ln {2^{2N}}}{N}=k_{B}\ln 4$ at half-filling.

In Fig. \ref{fig:entropy}, we present the entropy per particle of
the SU($4$) and SU($6$) Hubbard models as a function of $T$ at
various values of $U$.
In both cases, the $S(T)$ curves cross at a narrow region around a characteristic point ($T^{*}$, $S^{*}$). The characteristic
specific entropy $S^{*}$ increases with the number of fermion components
$2N$ as shown in Tab. \ref{crossing} summarized from the QMC results
of this paper and previous publications.
(On the square lattice, the Pomeranchuk effect is absent for the SU($2$)
fermions, because strong AF correlations of SU($2$) fermions
reduce the entropy capacity, while the multi-components of a large spin
suppress the AF correlations.)
Additionally $S^{*}$ is insensitive to the lattice structure and
the associated band structure of SU($2N$) fermions.
In fact, $S^*$ denotes the specific entropy of each particle coming
from the spin channel, which can be estimated as
\bea
S^*\approx \frac{1}{N}\ln \frac{(2N)!}{N! N!}.
\eea
In the SU(2), SU(4), and SU(6) cases, $S^*$'s are approximately
$\ln 2\approx 0.69$, $\frac{1}{2}\ln 6\approx 0.89$, and
$\frac{1}{3}\ln 20\approx 1.0$, respectively, which
excellently agrees with Tab. \ref{crossing}.

\begin{table}
\begin{ruledtabular}
\begin{tabular}{lll}
Symmetry & Lattice Type &  $S^{*}$  \\
\hline
SU$(2)$ & Square &          N/A    \cite{Paiva2010}    \\
   & Honeycomb &  $\sim0.65$  \cite{Tang2013}  \\
\hline
SU$(4)$ & Square &           $\sim0.9$  \cite{Zhou2014}   \\
   & Honeycomb &    $\sim0.9$  \\
\hline
SU$(6)$ & Square &           $\sim1.0$   \cite{Zhou2014} \\
   & Honeycomb &   $\sim1.0$ \\
\end{tabular}
\end{ruledtabular}
\caption{The characteristic specific entropy $S^{*}$ for spin components $2N=2,4,6$ with different lattice types.}
\label{crossing}
\end{table}

We first consider the low specific entropy regime $S<S^*$ in which the entropy
per particle increases monotonically with $U$ at a fixed
temperature.
At weak coupling $U$, the system is typically in the semi-metal state.
Its entropy is mainly contributed by fermions near the Dirac points,
and thus is small due to the vanishing of density of states.
As $U$ increases, the system becomes a Mott insulator, and
the on-site particle number fluctuations are still frozen in this
temperature regime.
As a result, fermions on each site contribute to the entropy by
means of spin fluctuations.
Hence, the semi-metal liquid-like phase is more ordered than the
solid-like state at the same temperature in the low temperature regime.
This is an example of the Pomeranchuk effect, which is first proposed
in the $^{3}$He system, where increasing pressure can further cool the
system in low temperature regime. The characteristic specific entropy
$S^*$ indicates the largest specific entropy for
exhibiting the Pomeranchuk effect, above which this effect disappears.
The Pomeranchuk effect was found more prominent in the SU($2N$) case
due to the enhanced entropy contribution from the spin channel
\cite{Hazzard2012,Cai2013,Zhou2014,Bonnes2012, Messio2012}.
On the honeycomb lattice, the density of states at weak $U$
is further suppressed in the semi-metal phase, and the AF
correlations are also weakened by the small coordination number.
As a result, the Pomeranchuk effect is more prominent than that
in the square lattice.

As shown in Fig. \ref{fig:entropy}, the fermion system can be driven to lower
temperatures by increasing the Hubbard $U$ adiabatically.
Particularly interesting, the isoentropy curve of $S/k_B=0.1$ intersects
the phase boundary near $(U/t=13, T/t=0.1)$ as shown in Fig. \ref{fig:phase},
which suggests a possible scenario for the experimental realization of
an SU(6) Mott insulator with the cVBS order.
In ultra-cold atom experiments, the interaction-induced cooling has
been achieved in optical lattices by fine tuning the Hubbard $U$
via Feshbach resonances \cite{Bloch2008}.

In the high specific entropy regime $S>S^{*}$, the entropy from the
spin channel has been fully used up.
Nevertheless, the contribution from the charge channel, {\it i.e.},
the fluctuations of the onsite particle number, becomes significant.
Increasing $U$ leads to the localization of fermions and thus suppresses charge
fluctuations.
As a result, the entropy per particle decreases with increasing $U$ at a fixed temperature in this
specific entropy regime.

The temperature regime for exhibiting the Pomeranchuk effect also
has a lower boundary.
On the honeycomb lattice, as shown in Fig.\ref{fig:entropy}
the Pomeranchuk effect becomes pronounced roughly starting
at $T/t \sim 0.1$ which is at the same temperature scale of the cVBS
ordering.
Below this temperature, the cVBS order develops, which dramatically decreases the entropy and then suppresses
the Pomeranchuk effect.

It is interesting to note that, similar to the narrow crossing of entropy curves revealed in our simulation, the narrow crossing of specific heat curves was studied by Vollhardt in spin-$1/2$ correlated systems \cite{Vollhardt1997}. Following the same reasoning, in the next section we shall explain analytically the narrowness of crossing region of entropy curves in the SU($2N$) case.

 \subsection{The narrow crossing of entropy curves}
 \label{narrowST}
\begin{figure}[htb]
 \includegraphics[width=0.9\linewidth]{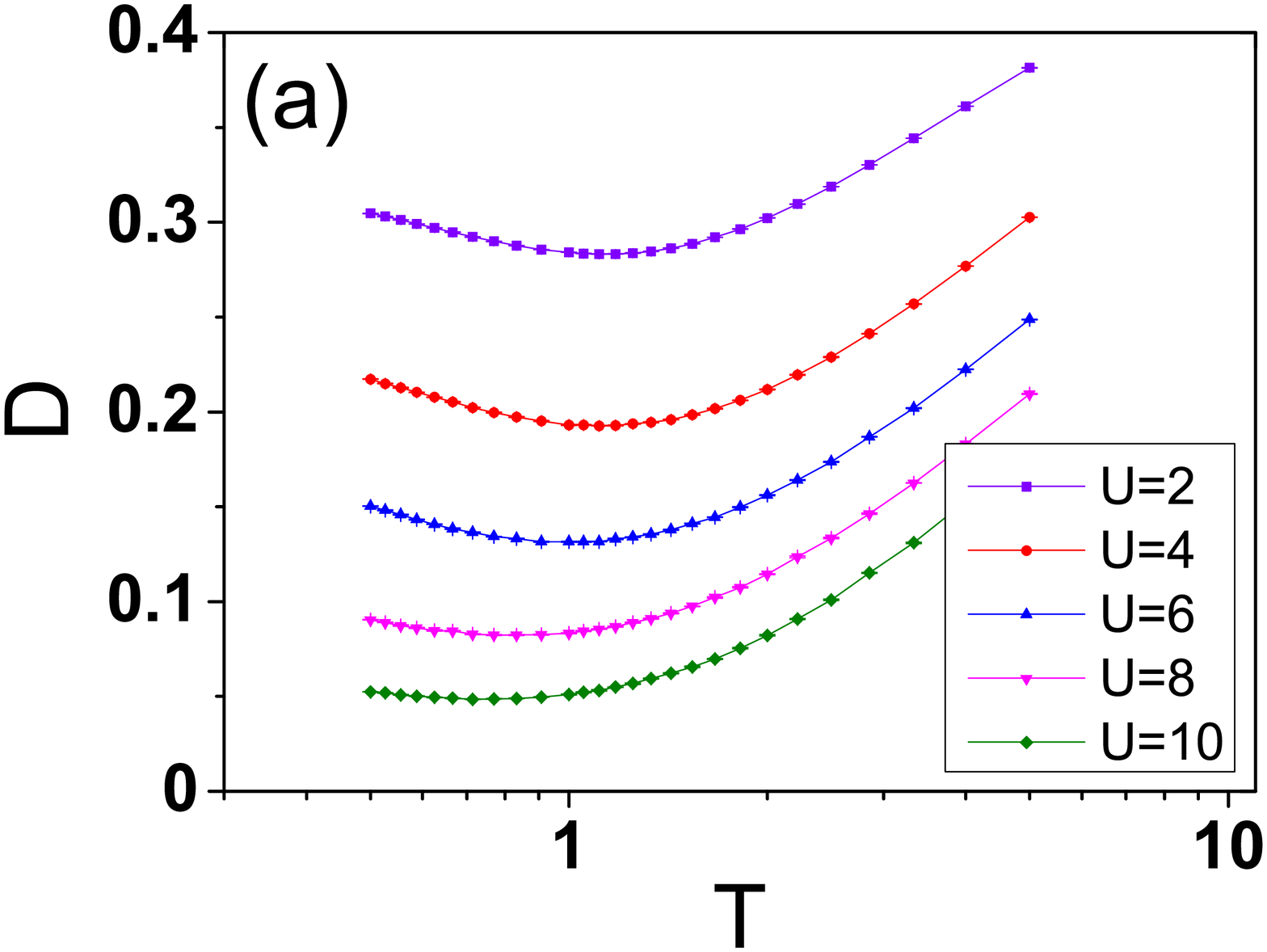}
 \includegraphics[width=0.9\linewidth]{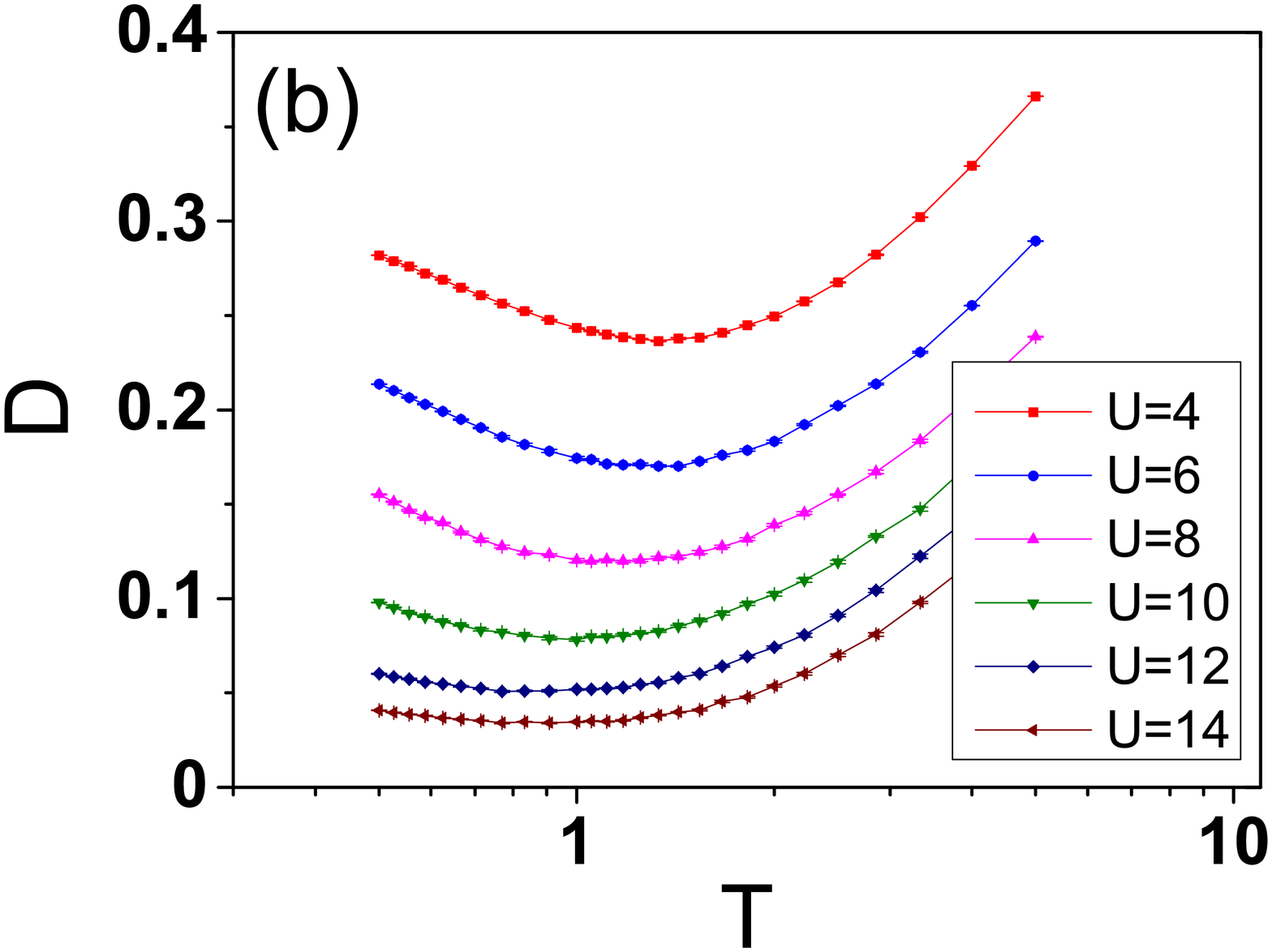}
 \caption{$D$ as a function of temperature $T$ for ($a$) SU($4$) and ($b$) SU($6$)
  Hubbard models at half filling.
  The system size is $L=9$.
  Error bars are smaller than the data points.
 }
 \label{fig:TD}
 \end{figure}

 Along the same line as Vollhardt's work for the SU($2$) case \cite{Vollhardt1997}, in the SU($2N$) case the conjugate intensive variable associated with $U$ is
 \be
 D(T,U) = \frac{1}{2L^2}\frac{\partial F(T,U)}{\partial U } = \frac{1}{4L^2}\sum_{i\in A \oplus B}\left(n_{i}-\langle n_i\rangle\right)^{2},
 \ee
 where $F(T,U)$ is the free energy.
 At half-filling, the average particle number per site is $\langle n_i\rangle=N$,
 and $D$ serves as the variance characterizing the on-site particle number fluctuation.
Especially for the SU($2$) case, $D$ is just the on-site double occupancy\cite{Vollhardt1997, Werner2005, Jordens2008}.
The temperature dependence of $D$ is calculated for a range of Hubbard $U$, as shown in Fig. \ref{fig:TD}. It is seen that the on-site particle number fluctuations $D$ are suppressed with increasing $U$. The temperature dependence of $D$ is non-monotonic due to the Pomeranchuk effect. For each $U$,
the on-site particle number fluctuation $D$ achieves the minimum at around $T^{*}\sim t$.

The entropy per site $S$ and the on-site particle number fluctuation $D$ satisfy the Maxwell relation
 \be
 \frac{\partial S(T,U)}{\partial U}=-\frac{\partial D(T,U)}{\partial T}.
 \label{eq:maxwell}
 \ee

We first illuminate why the entropy curves cross. Note that the on-site particle number fluctuations $D$ reach their minima at around $T^{*}\sim t$ regardless of the coupling strength $U$, i.e., $\partial D(T^{*},U)/ \partial T^{*}$=$0$. Using the Maxwell relation (\ref{eq:maxwell}), one finds that $\partial S(T^{*},U)/ \partial U$=$0$, which implies the crossing of entropy curves at around $T^{*}\sim t$.

We now explain the narrowness of the crossing region. We expand $S(T^{*},U)$ as a power series in $U-U_{0}$, with $U_{0}$ chosen at convenience. To the leading term, one obtains
 \be
 S(T^{*},U) \approx S(T^{*},U_{0}) \left[1+\frac{1}{2}\frac{(U-U_{0})^2}{S(T^{*},U_{0})}\frac{\partial^2 S(T^{*},U)}{\partial U^2}\mid_{U=U_0} \right].
 \label{eq:expansion}
 \ee
 The width of the crossing region is then determined by the curvature of the entropy $S(T,U)$ at $T^{*}$. Using Eq. (\ref{eq:maxwell}) and $\partial D(T^{*},U)/ \partial T^{*}$=$0$ , one obtains
 \be
 \frac{\partial^2 S(T^{*},U)}{\partial U^2}  =- \frac{\partial}{\partial U} \left[ \frac{\partial D(T^{*},U)}{\partial T^{*}} \right] = 0.
 \ee
which guarantees the entropy curves cross at a narrow region around a characteristic point $(T^{*},S^{*})$.

\subsection{ The probability distributions of on-site occupation number}

\begin{figure}[htb]
\includegraphics[width=0.8\linewidth]{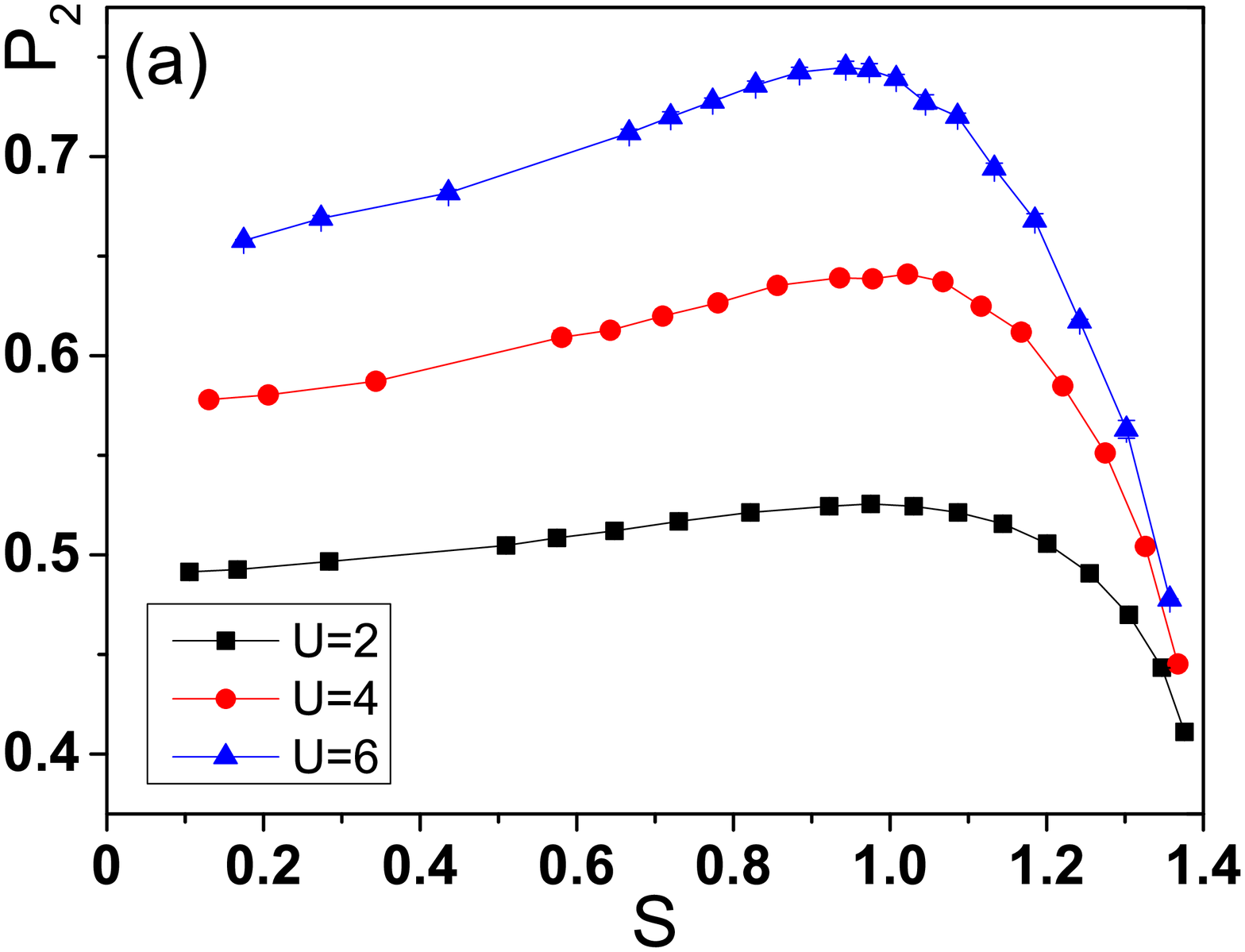}
\includegraphics[width=0.8\linewidth]{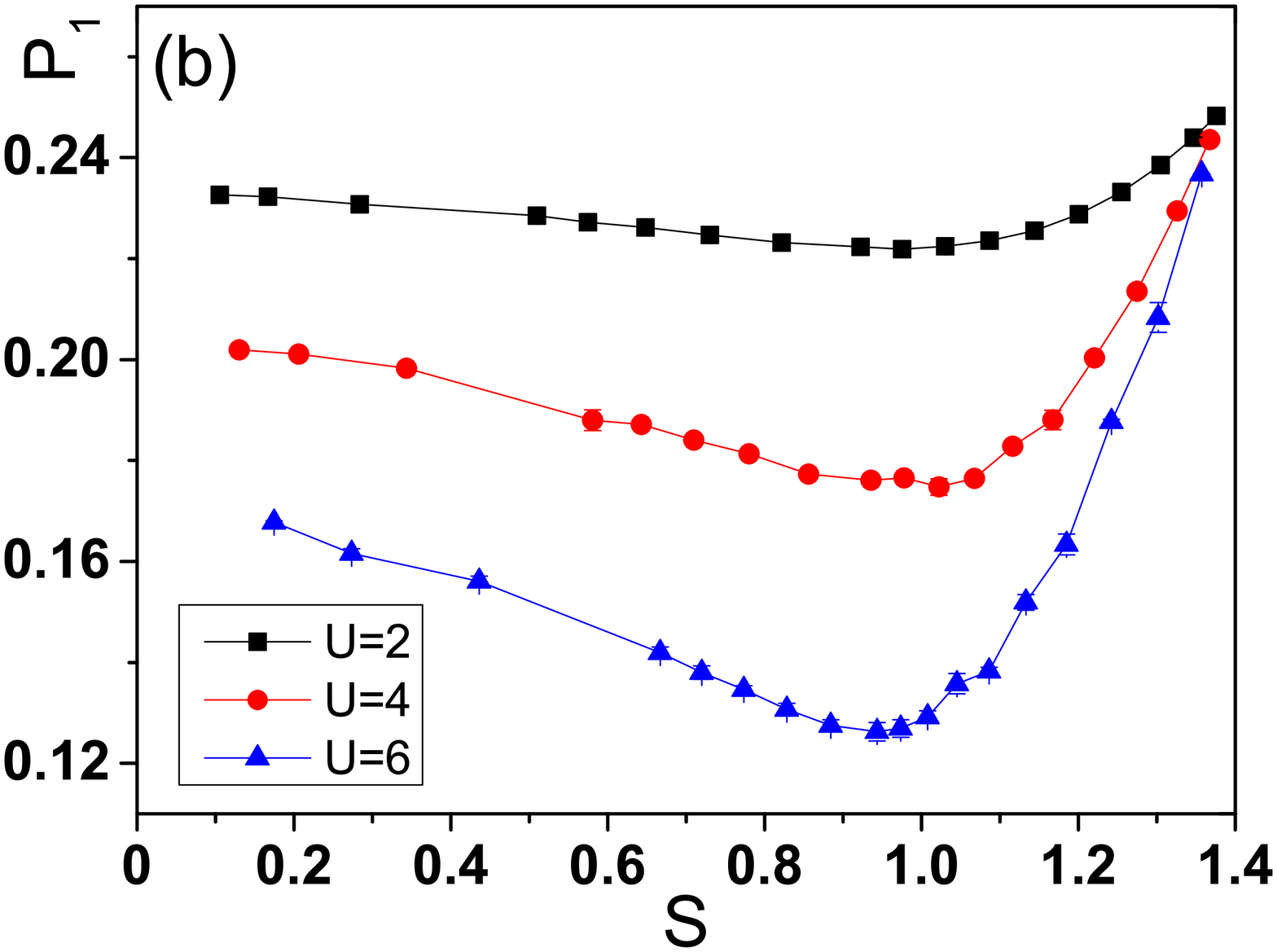}
\includegraphics[width=0.8\linewidth]{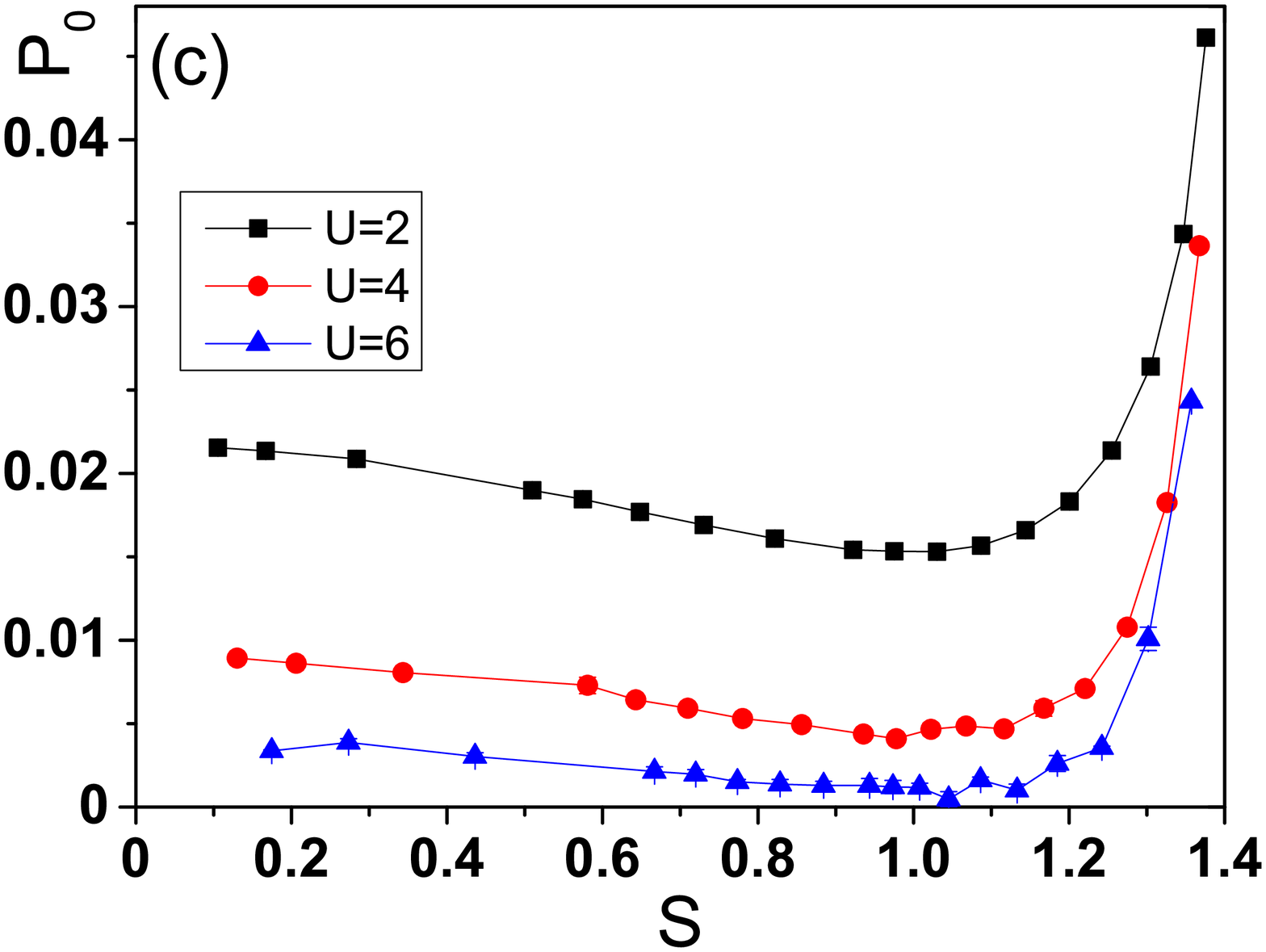}
\caption{The probability distributions $P(n)$ of the on-site particle numbers
($a$) $P(2)$, ($b$) $P(1)$, and ($c$) $P(0)$ versus entropy $S$
at different values of $U$ in the half-filled SU($4$)
Hubbard model.
The lattice size is $L=9$.
The lines serve as a guide to the eye, and error bars are smaller
than the data points.
}
\label{fig:moment}
\end{figure}

In the SU(2) Hubbard model, the double occupancy is a physical
observable in cold atom experiments \cite{Jordens2008,Strohmaier2010}.
This quantity behaves slightly non-monotonic with temperature \cite{Werner2005}.
In the high temperature regime where thermal fluctuations dominate,
the double occupancy can be used as thermometers \cite{Jordens2010}.
In this part, we will simulate the half-filled SU($4$) Hubbard model,
illustrating the relation between entropy and the distributions
of on-site particle numbers.

The probability distribution $P(n)$ of the on-site occupation number $n$
is defined as \cite{Zhou2014},
\bea
P(0)&=&\prod^{4}_{\alpha=1} (1-n_{i}^{\alpha}),
\nn \\
P(1)&=&\sum_{\alpha=1}^{4} n_{i}^{\alpha}\prod_{\beta\ne\alpha}(1-n_{i}^{\beta}),
\nn \\
P(2)&=&\sum_{\alpha\ne\beta}n^{\alpha}_{i}n^{\beta}_{i}
\prod_{\gamma\ne\alpha\beta}(1-n_{i}^{\gamma}).
\eea
where $n_{i}^{\alpha}$ is the particle number operator on site $i$ with
spin $\alpha$.
The total probability is normalized to unity.
The particle number fluctuations also obey the particle-hole symmetry at
half-filling and thus $P(0)=P(4)$ and $P(1)=P(3)$.
In the high temperature limit, the on-site occupation number obeys
the binomial distribution, which is, $\lim_{T\rightarrow \infty} P(k)
= C^{k}_{4}/ 2^{4}$.

In Fig. \ref{fig:moment}, the relationships between $P(n)$ with $n=0,1,2$
and the entropy $S$ are presented.
The distributions are dramatically non-monotonic with entropy even in
the weak coupling regime, e.g. $U/t=2,4$, and $6$.
As expected, at half-filling, the most probable distribution of the SU($4$)
Dirac fermions on each site is the double occupancy, and the deviation
from the double occupancy is due to charge fluctuations.
These curves show that roughly when $S<S^*\approx 0.89$, the onsite charge
fluctuations decrease with increasing entropy, and fermions tend to
localize in consistent with the Pomeranchuk effect.
This seemingly discrepancy is due to the dominant entropy contribution
from the spin channel.
In cold atom experiments, the site-resolved quantum gas microscopy can be
used to detect the on-site particle number distributions \cite{Preiss2015}.

\section{The density and spin responses}
\label{response}
In this section, we investigate the density compressibility and the
uniform spin susceptibilities of the half-filled SU($2N$) Hubbard
model on a honeycomb lattice.

\subsection{The density compressibility}
The density compressibility is defined as
\bea
\kappa = \frac{\beta}{2L^2} \left( \langle (\sum_{i}n_i )^2\rangle
- \langle \sum_{i}n_i \rangle ^2   \right),
\eea
which is related to the global density fluctuations.
It is an observable in cold atom experiments.
The vanishing of $\kappa$ at low temperatures is a characteristic signature
of the Mott insulating states \cite{Schneider2008, Duarte2015}.

\begin{figure}[htb]
\includegraphics[width=0.9\linewidth]{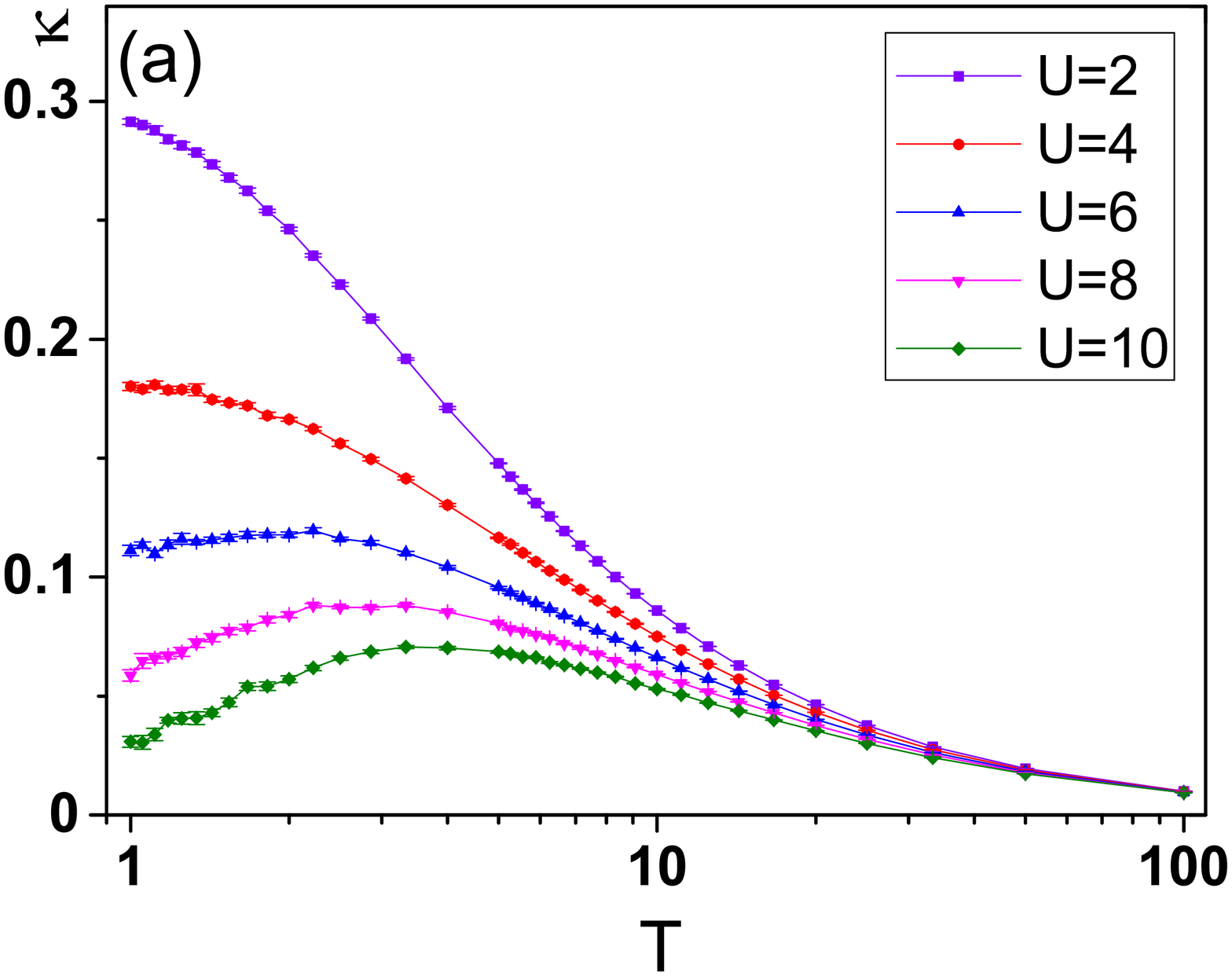}
\includegraphics[width=0.9\linewidth]{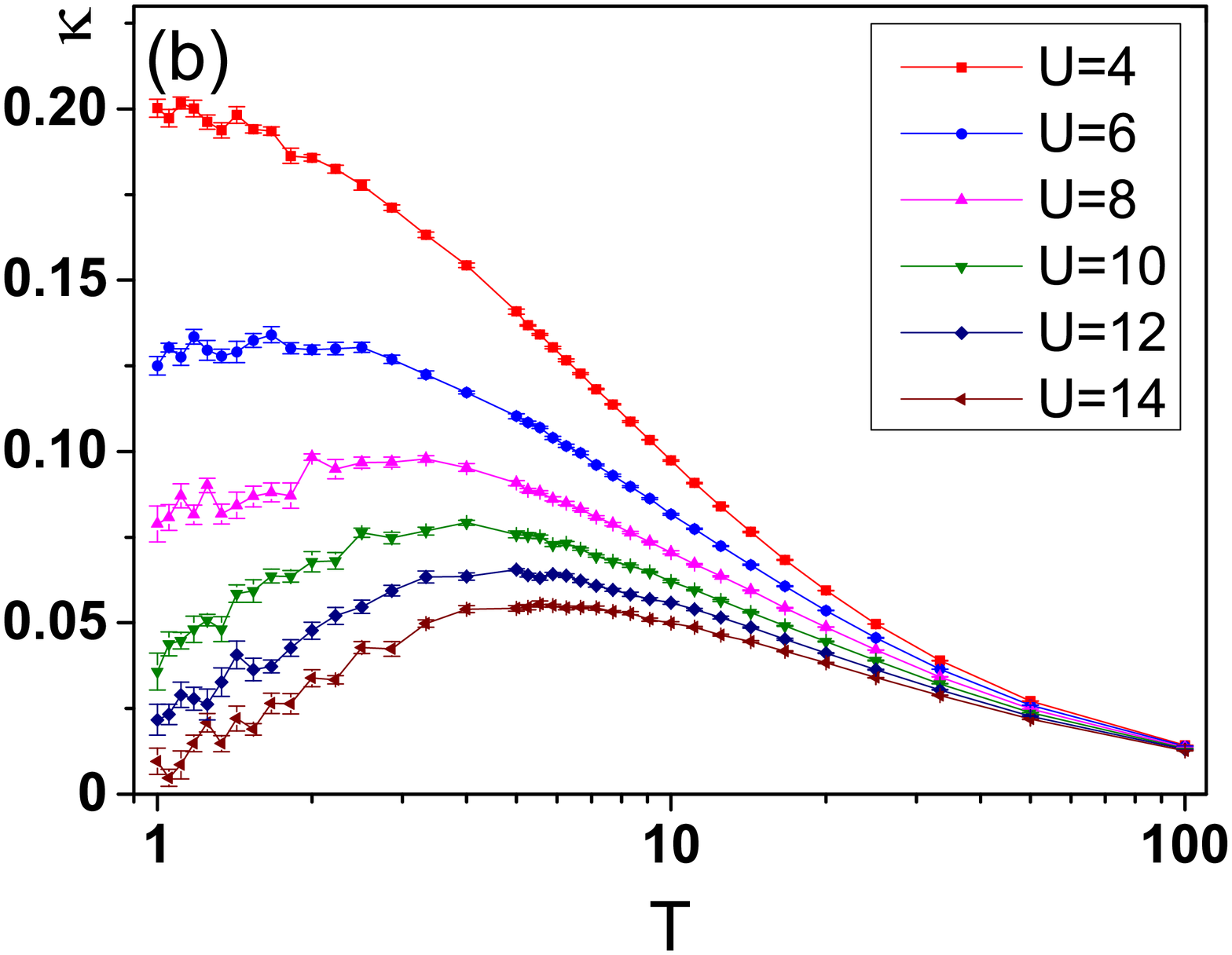}
\caption{The density compressibility $\kappa$ versus $T$ at
various values of $U$ in the half-filled ($a$) SU(4) and ($b$) SU(6)
Hubbard models. The lattice size is $L=9$.
}
\label{fig:com}
\end{figure}

We present the DQMC simulation results for the density compressibility
of SU(4) and SU(6) Hubbard models on a honeycomb lattice in
Fig. \ref{fig:com} ($a$) and ($b$), respectively.
Here we only calculate $\kappa$ in the temperature regime corresponding to $S>S^*$ segment of the $S(T)$ curves.
At low temperatures, the simulation of $\kappa$ becomes numerically
unstable as explained in Appendix \ref{app:com}.
At very high temperatures $T \gg  U$, $\kappa(T)$ behaves like that of
a classic ideal gas, i.e., $\kappa \sim 1/T$, which means charge
incoherence.
On the other hand, increasing $U$ while fixing $T$ suppresses
the compressibility.
However, in the zero temperature limit not shown in Fig. \ref{fig:com},
$\kappa$ should go to zero both in the Dirac semi-metal phase due
to the vanishing of density of states, and in the cVBS state due to
the charge gap opening.
Consequently, the $1/T$ divergence of $\kappa$ stops when $T$
decreases to a certain temperature scale dependent of $t$ and $U$.
At large values of $U>U_c$ where $U_c$ is the
critical interaction strength for the emergence of the cVBS ground state,
$\kappa$ becomes decreasing along with lowering $T$ after reaching
the maximal value at a temperature comparable to $U$.
($U_c\approx 7$ and $11$ in the SU(4) and SU(6) cases,
respectively \cite{Zhou2016}.)

Note that, in the SU(6) case with $U/t = 14$, $\kappa$ is nearly
suppressed to zero at $T/t \sim 1$, a temperature scale comparable
to the band width but still much smaller than the Hubbard interaction.
This is also the temperature scale for the thermal melting of
the cVBS state as shown in Fig. \ref{fig:phase}.
Thus the finite-temperature disordered states outside the cVBS phase
exhibit different characters: in the weak coupling side, it is
a finite-temperature semi-metal state, while in the strong
coupling side, it is a finite-temperature Mott-insulating
state with thermally melted cVBS order.
Nevertheless, they can be smoothly connected at finite temperatures.

\begin{figure}[htb]
\includegraphics[width=0.9\linewidth]{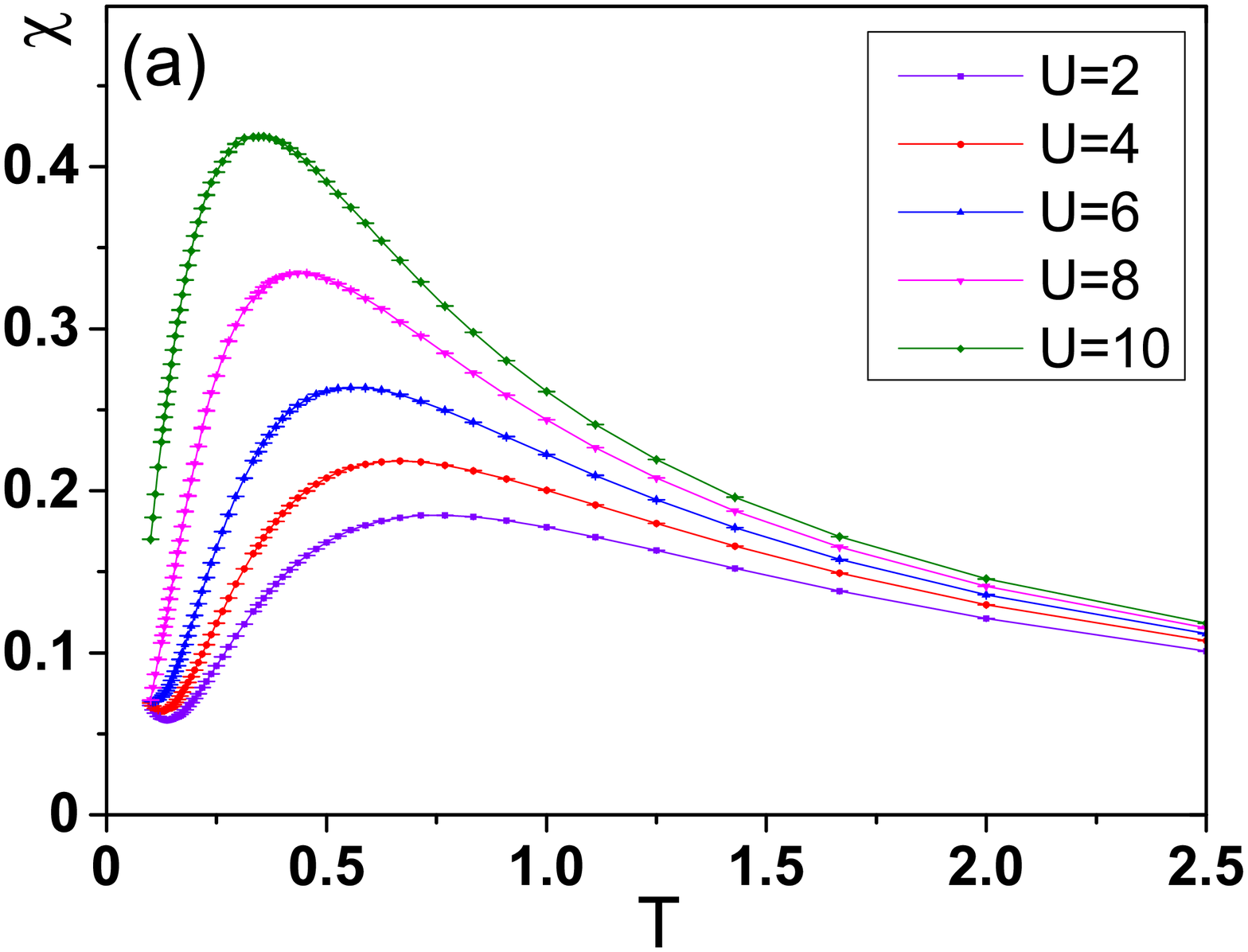}
\includegraphics[width=0.9\linewidth]{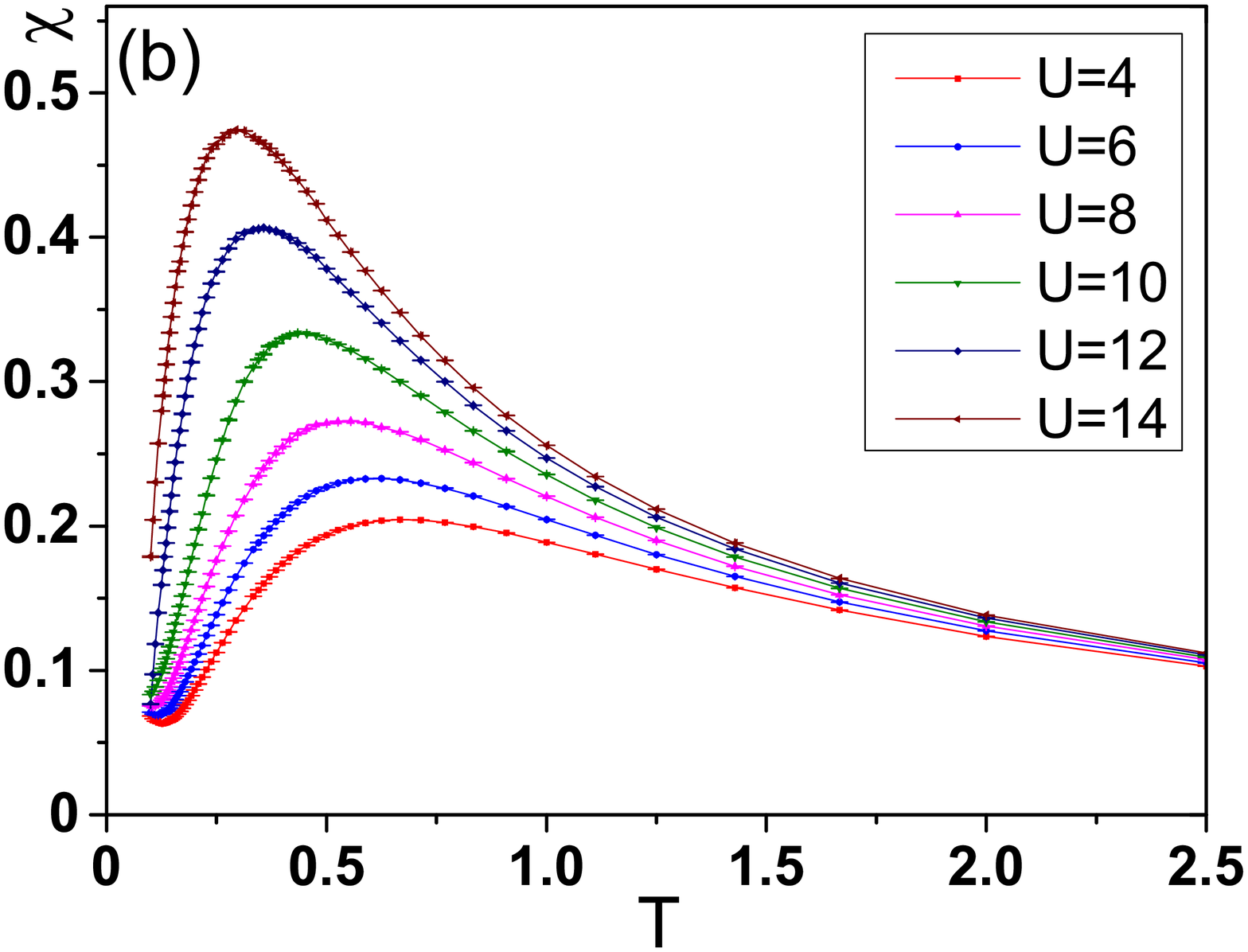}
\caption{The uniform spin susceptibilities $\chi$ versus
temperature $T$ at various values of $U$ in the half-filled
($a$) SU($4$) and ($b$) SU($6$) Hubbard models.
The lattice size is $L=9$. Error bars are smaller than
the data points.
}
\label{fig:TX}
\end{figure}

\subsection{The uniform spin susceptibilities}

The uniform spin susceptibility $\chi$ is defined as
\bea
\chi = \frac{\beta}{2L^2}\sum_{i,j}S_{spin}(i,j),
\eea
where $S_{spin}(i,j)$ is the SU($2N$) version of the equal-time
spin-spin correlation:
\bea
S_{spin}(i,j)=\frac{1}{(2N)^2-1}\sum_{\alpha,\beta}
\langle S_{\alpha\beta,i}S_{\beta\alpha,j}\rangle.
\eea
Note that
$S_{\alpha\beta,i}=c^{\dagger}_{\alpha,i}c_{\beta,i}-\frac{\delta^{\alpha\beta}}{2N}
\sum_{\gamma=1}^{2N} c_{\gamma,i}^{\dagger}c_{\gamma,i}$
are the generators of an SU($2N$) group and satisfy
the commutation relation
$[S_{\alpha\beta},S_{\gamma\delta}]=\delta^{\beta\gamma}
S_{\alpha\delta}-\delta^{\alpha\delta} S_{\gamma\beta}$.

In Fig. \ref{fig:TX}, the uniform spin susceptibility $\chi(T)$'s of
the SU($4$) and SU($6$) Hubbard models on a honeycomb lattice
are plotted for various values of Hubbard $U$.
The high temperature behaviors of $\chi(T) \sim 1/T$ obey the
Curie-Weiss law, which shows the spin incoherence.
Again this divergence is suppressed at low temperatures since in the
zero temperature limit, $\chi(T)$ approaches zero in both the
Dirac semi-metal phase and the cVBS phase.
In the former case, it is because of the vanishing of density of states,
while in the latter case, it is due to that the cVBS phase is a spin gapped phase.
Thus a peak in each $\chi(T)$ curve must develop in the full range of the
Hubbard $U$.
In the strong coupling regime, the peak is located around the super-exchange
energy scale $J\approx 4t^2/U$.
In contrast, the peak location in the weak coupling regime is mostly
determined by the band width $t$, and consequently the peak is located at the energy
scale in which the density of states becomes linear.

One observation from Fig. \ref{fig:TX} is that $\chi(T)$
increases monotonically with $U$ at a fixed temperature.
In the weak coupling regime, this is consistent with the mean-field
analysis concluding that the uniform spin susceptibility $\chi$
is enhanced by the repulsive interaction \cite{Fazekas1999}.
In the strong coupling regime, increasing $U$ enhances the amplitudes
of the onsite spin moments by suppressing the change fluctuations,
and thus $\chi(T)$ is also increased.
At small values of $U$, a tiny upturn occurs in the $\chi(T)$ curve
at low temperatures, which is caused by the finite-size effect
\footnote{In $U=0$ case, the spin susceptibilities are calculated by
$\chi(T)=- \int_{-\infty}^{\infty}d\epsilon \frac{\partial f}
  {\partial \epsilon} \rho(\epsilon)$, where $f$ is the Fermi-Dirac
  distribution and $\rho$ is the density of states.
 In finite-size system, the integration has been transformed into summations.
 So the discrete energy levels will take the responsibility of the upturn.
 In Ref.[\onlinecite{Peres2006}], a small upturn is
 also reported in $\chi(T)$.  }.

Considering the SU(4) and SU(6) Hubbard models on a honeycomb lattice,
the AF ordering does not occur even in the ground state.
Nevertheless, we also present the simulation results of the AF structure
factors and nearest-neighbor spin-spin correlations in Appendix
\ref{app:AF}.

\section{Conclusions}
\label{ConDiscu}

In summary, we have employed the large-scale DQMC simulations to study
the effects of SU($2N$) symmetries on thermodynamic properties of
Dirac fermions.
The Dirac fermions are described in terms of the SU($2N$) Hubbard model
on a honeycomb lattice which captures the interplay between charge
and spin degrees of freedom.
We have simulated the finite-temperature properties of SU(4) and SU(6)
cases, including the thermal VBS phase transition, the Pomeranchuk effect,
the density compressibility, and spin susceptibilities.

We use the SU(6) case as an example to study the thermal phase transition
between the disordered state and the cVBS state on a honeycomb lattice.
In the SU(2) honeycomb-lattice Hubbard model, the Mott insulating phase
at $T=0$ exhibits the AF ordering which breaks the continuous SU(2)
symmetry and thus cannot exist at finite temperatures.
Nevertheless, the cVBS order in the SU(6) case only breaks a discrete
symmetry and does occur in the thermal transition.
Based on the above reasoning, the thermal cVBS phase transition
is also expected in the simulations of the SU(4) honeycomb-lattice
Hubbard model, though the cVBS order is weaker compared
with the SU(6) case.
The simulation of entropy-temperature relations shows that the $S(T)$
curves with different Hubbard $U$ cross at a narrow region around a characteristic point ($T^{*}$, $S^{*}$)
characterizing the onset of the Pomeranchuk effect.
This characteristic specific entropy $S^{*}$ comes from the local spin moment contribution
estimated as $S^*\approx\frac{1}{N}\ln \frac{(2N)!}{N!N!}$.
As demonstrated in our DQMC simulations, the SU($6$) cVBS Mott insulating state can be reached along the isoentropy
curve $S/k_B=0.1$ by the interaction-induced adiabatic cooling, which sheds
new light on future explorations of novel states of matter
with ultra-cold $^{173}$Yb  experiments.

It is worth noting that, a plateau of $S=S^{*}$ is expected to appear in a single $S(T)$ curve when Hubbard
$U$ is large enough, due to the full release of spin entropy. In fact, the roles of spin and charge channels in entropy production are separated at around $S=S^{*}$,
when the Coulomb repulsion $U$ becomes stronger than the critical interaction that leads to the emergence of the cVBS ground state.
Interestingly, in the weak and intermediate coupling regimes, the simulated $S(T)$ curves cross at around a characteristic point where $S=S^{*}$, though $S^{*}$ is not noticeable in a single $S(T)$ curve. The underlying physics of this special phenomenon may be revealed in future studies.

\acknowledgments
Z. Z. and Y. W.  gratefully acknowledge financial support from the National
Natural Science Foundation of China under Grant Nos. 11574238 and 11328403. D. W. acknowledges the support from National Natural Science
Foundation of China (11504164).
C. W. is supported by the NSF DMR-1410375 and AFOSR FA9550-14-1-0168.
C. W. acknowledges the support from the Presidents Research Catalyst
Awards of University of California.
This work made use of the facilities of Tianhe-2 at the China's
National Supercomputing Centre in Guangzhou.

\appendix

\section{The finite-size effect on entropy}
 \label{app:U0_entropy}

 \begin{figure}[htb]
 \includegraphics[width=0.8\linewidth]{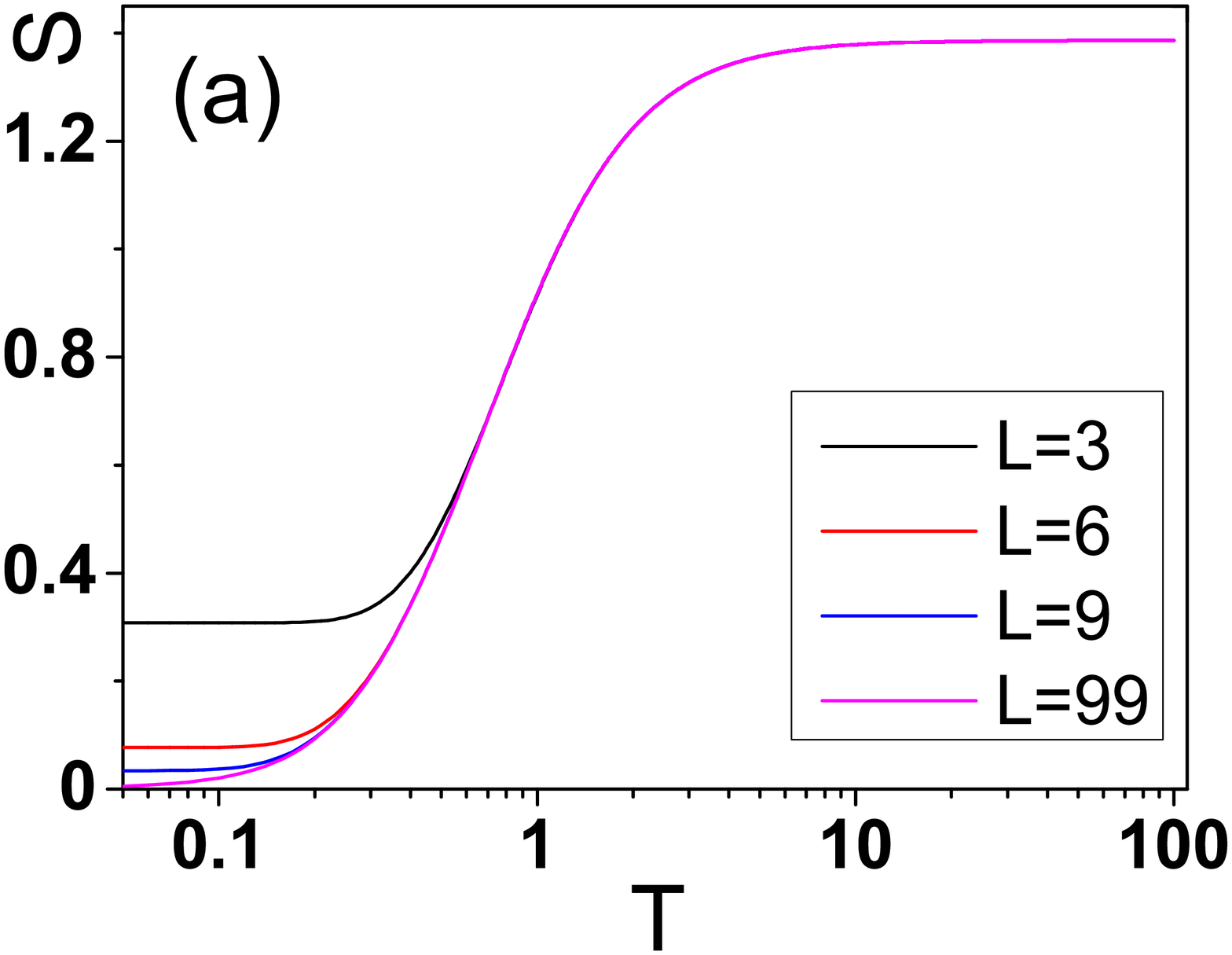}
 \includegraphics[width=0.8\linewidth]{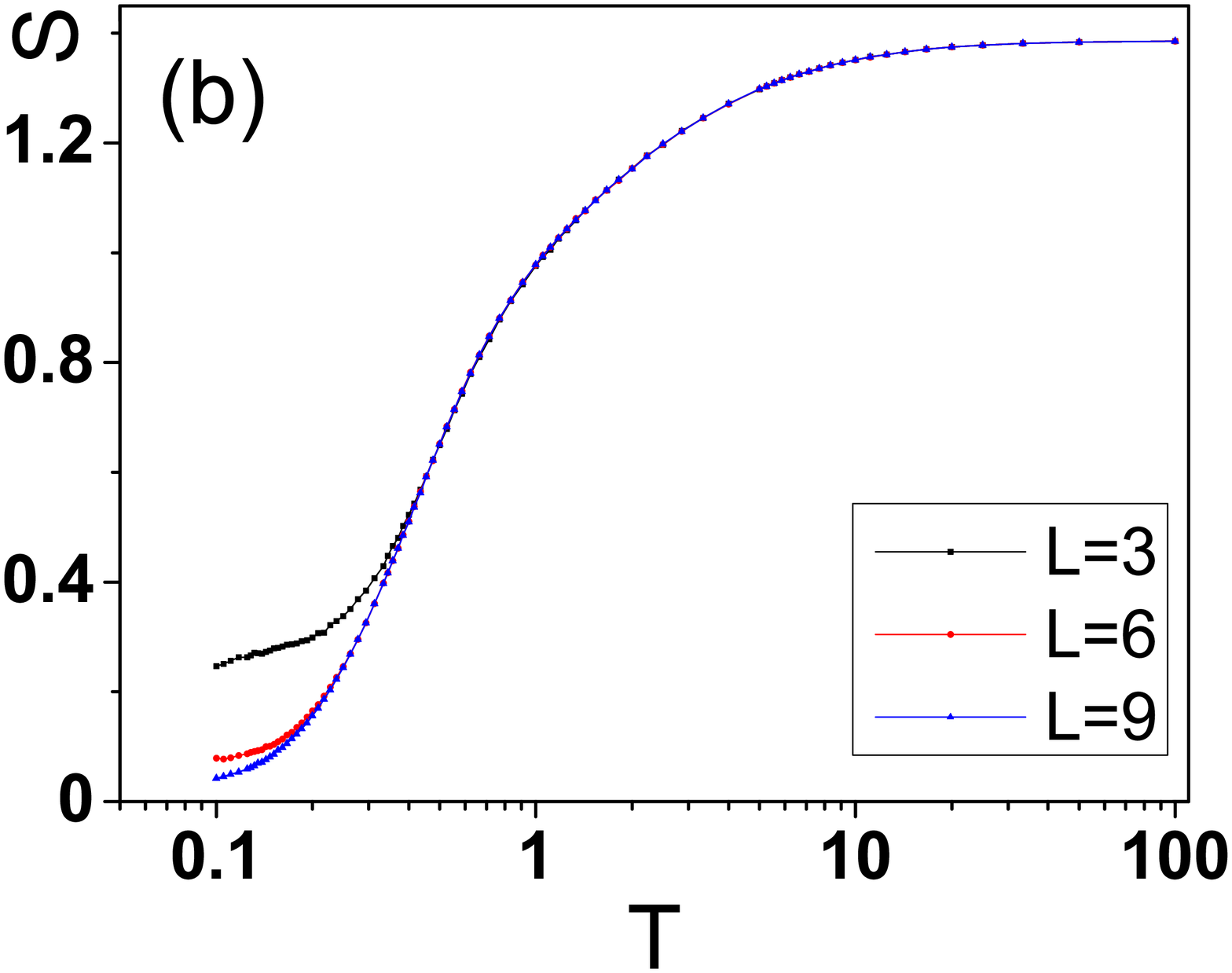}
 \includegraphics[width=0.8\linewidth]{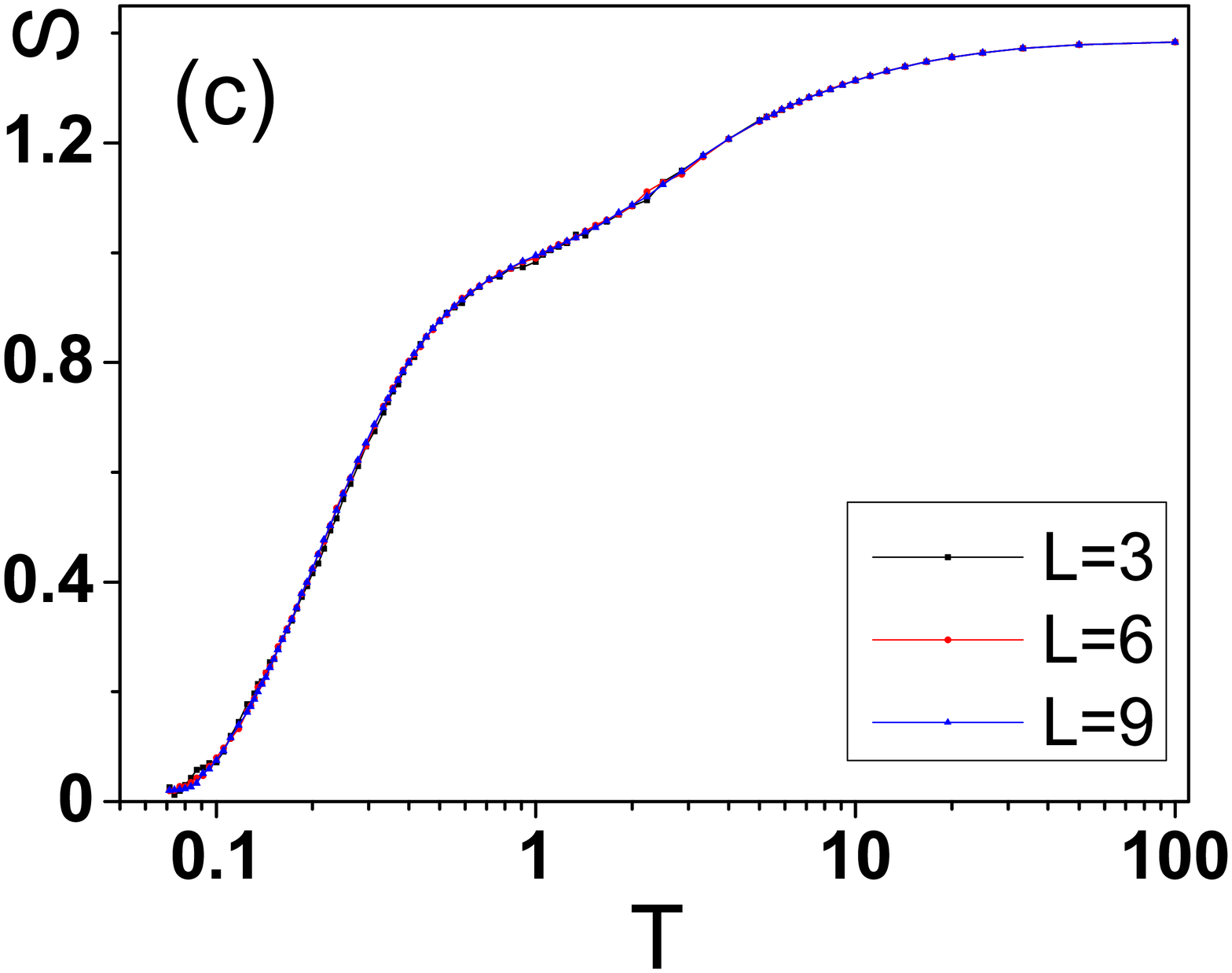}
 \caption{The finite-size dependence of entropy per particle of the half-filled SU(6) Hubbard model
  with parameters ($a$) $U=0$,
 ($b$) $U=6$, and ($c$) $U=12$.
 }
 \label{fig:FSE_ent}
  \end{figure}

 In the weak coupling regime, the finite-size effect is significant at low temperatures.
 We investigate the finite-size dependence of entropy per particle of a half-filled SU($2N$) tight-binding model
 on a honeycomb lattice.
 The entropy per particle can be calculated by \cite{Dare2007}:
 \be
 S(T,U=0) = -\frac{1}{L^{2}}\sum_{k} (f\ln f + (1-f)\ln(1-f)),
 \ee
 where $f$ is the Fermi-Dirac distribution.
As shown in Fig. \ref{fig:FSE_ent} ($a$), the residue entropy caused by finite-size effect decreases with increasing lattice size.
It is seen that the finite-size effect is not severe for  $L=9$ with $S/k_B \ge 0.1$.

We can see in Fig. \ref{fig:FSE_ent} ($b$) that the finite-size effect still exists for $U=6$ in the semi-metal region. But the dimer formation in a bond lifts degeneracy and thus lower the entropy in the strong coupling regime, as shown in Fig. \ref{fig:FSE_ent} ($c$) where $U=12$.
 Moreover, the cVBS correlation length is much larger than the lattice size.
As a result, the finite-size effect is weak in the cVBS region.
Hence the isoentropy curve demonstrating Pomeranchuk cooling in Fig. \ref{fig:phase} is a reasonable estimate in the thermodynamic limit.

\section{Imaginary part of the compressibility}
\label{app:com}

\begin{figure}[htb]
\begin{minipage}{4.2cm}
\includegraphics[width=4.9cm]{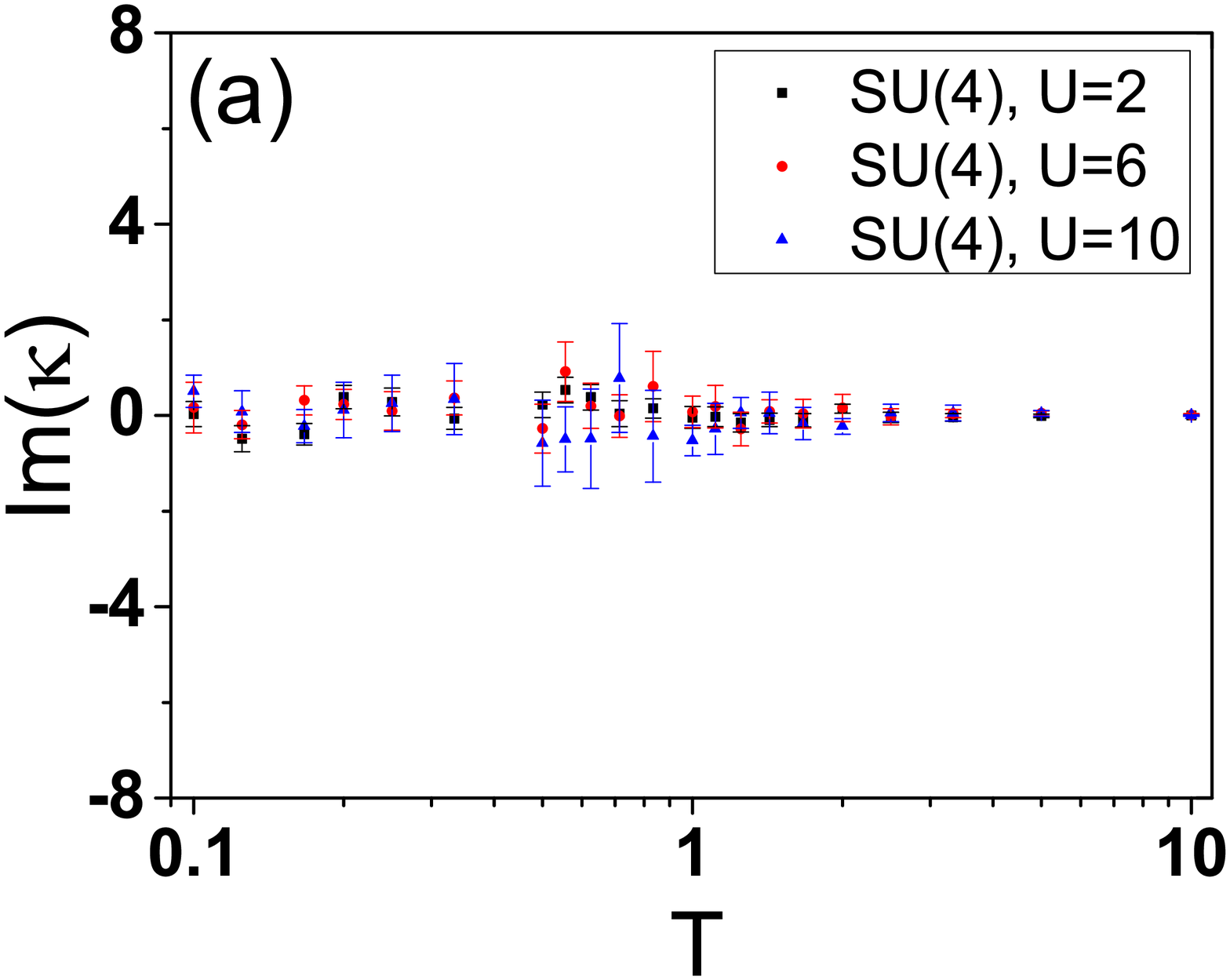}
\end{minipage}
\begin{minipage}{4.2cm}
\includegraphics[width=4.9cm]{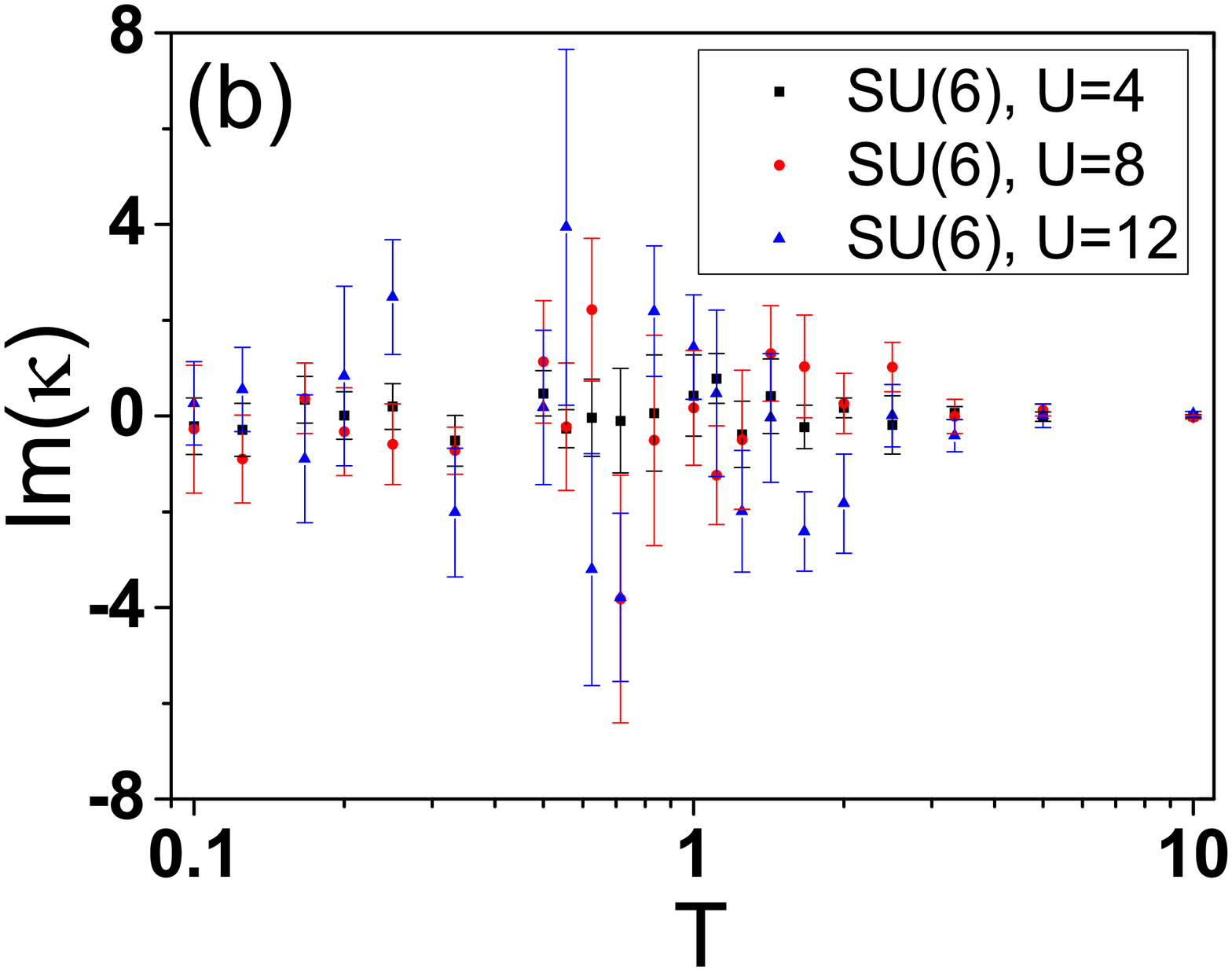}
\end{minipage}

\begin{minipage}{4.2cm}
\includegraphics[width=4.9cm]{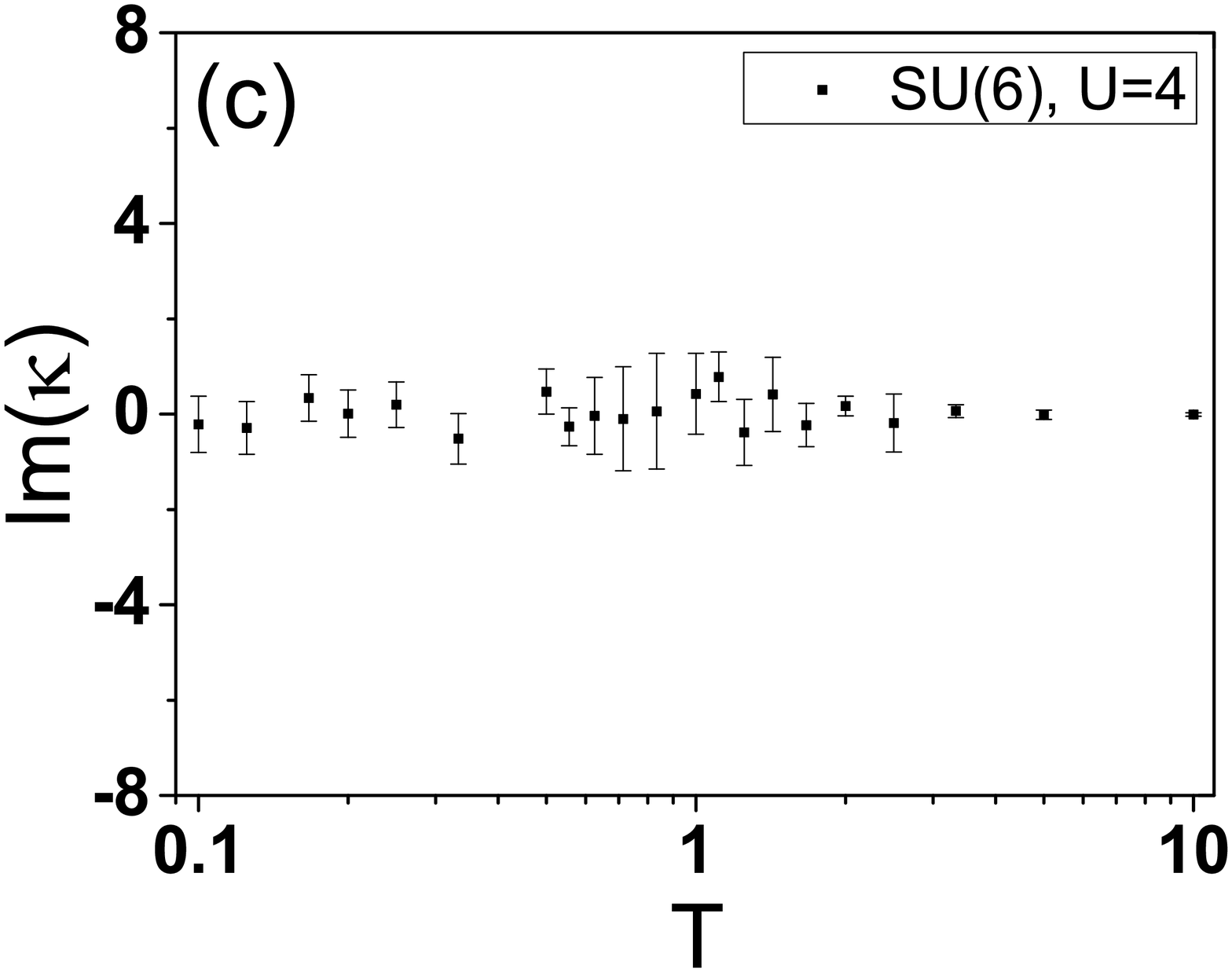}
\end{minipage}
\begin{minipage}{4.2cm}
\includegraphics[width=4.9cm]{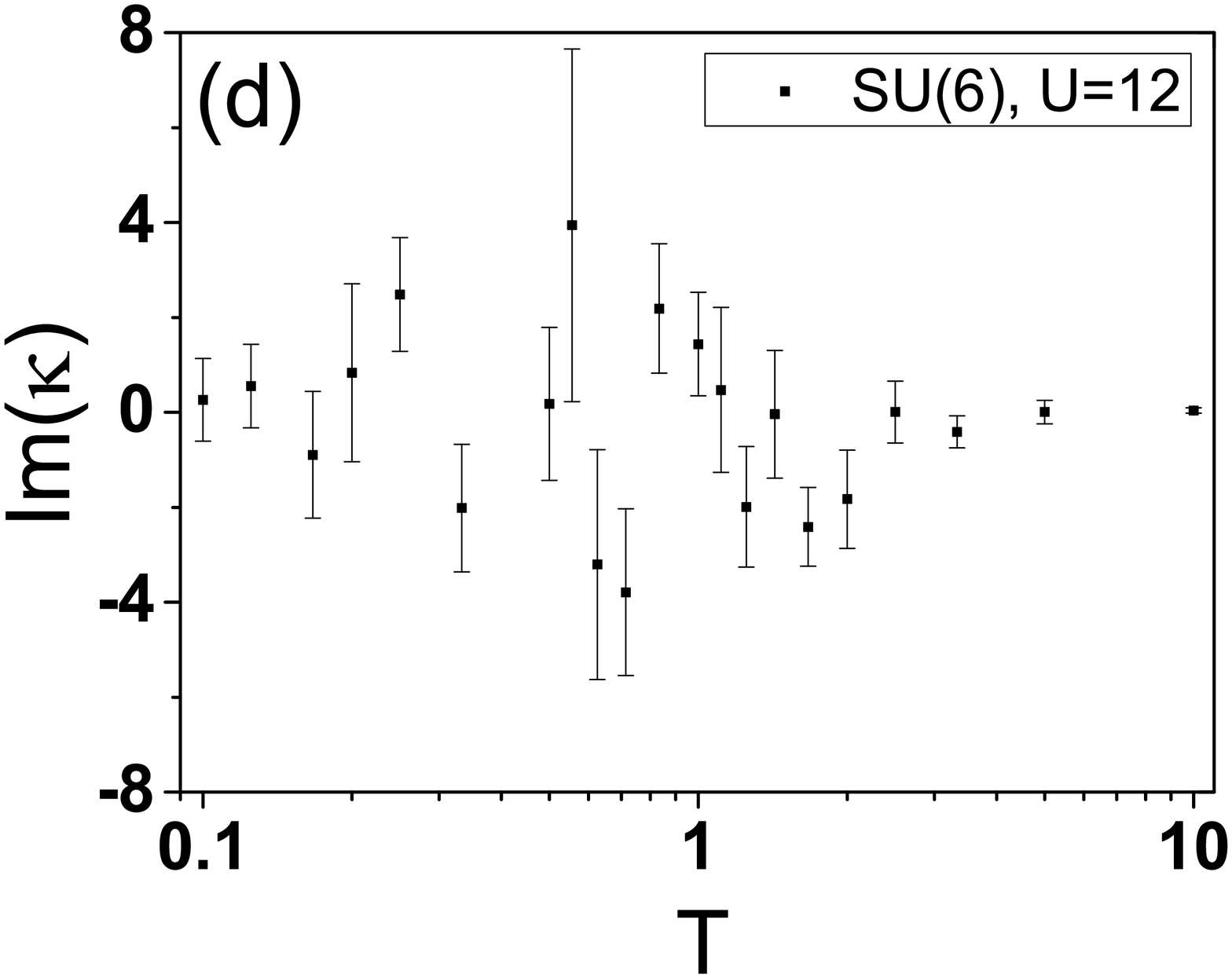}
\end{minipage}
\caption{The imaginary part of compressibility $\Im (\kappa)$ versus $T$
with different $U$ and $2N$: ($a$) SU($4$) case with different $U$;
($b$) SU($6$) case with different $U$;
($c$) SU($6$) case with $U=4$;
($d$) SU($6$) case with $U=12$.
}
\label{fig:appcom}
\end{figure}

In our simulations, the Hubbard-Stratonovich transformation is
performed in the density channel as below,
\bea
e^{\frac{\Delta \tau U}{2} (n_{j}-N)^{2} } = \frac{1}{4} \sum_{l=\pm 1}
\gamma _{j}(l)e^{i\eta_{j}(l)(n_{j}-N)}
\eea
where $\gamma$ and $\eta$ are two sets of parameters.
According to Ref. [\onlinecite{Wang2014}], in the cases of $2N=2, 4$ and $6$, the Ising fields can take values of
\begin{eqnarray}
\nonumber \gamma(\pm1)&=&\frac{-a(3+a^{2})+d}{d},\\
\nonumber \gamma(\pm2)&=&\frac{a(3+a^{2})+d}{d},\\
\nonumber \eta(\pm1)&=&\pm\cos^{-1}
\left\{ \frac{a+2a^{3}+a^{5}+(a^{2}-1)d}{4}\right\},\\
\nonumber \eta(\pm2)&=&\pm\cos^{-1}
\left\{ \frac{a+2a^{3}+a^{5}-(a^{2}-1)d}{4}\right\},
\end{eqnarray}
where $a=e^{-\Delta\tau U/2}$, and $d=\sqrt{8+a^{2}(3+a^{2})^{2}}$.

Because the diagonal term is complex, the decomposed fermion bilinear
operators are no longer Hermitian.
If all the configurations are reached when performing the path integrals,
the Hermitian of the many-body Hamiltonian is recovered.
However, since the importance sampling is used in the Monte Carlo
integrations, the imaginary part of a physical quantity is
only statistically zero.

The compressibility $\kappa(T)$ is related to the global density-density
correlations rather than local on-site correlations.
In Fig. \ref{fig:appcom}, we calculate the imaginary part of $\kappa(T)$
with different values of $T$, $U$, and $2N$.
$\Im(\kappa)$ fluctuates around zero severely in the low temperature regime.
Furthermore, it is seen that fluctuations turn to be increasingly
severe when increasing the value of $2N$ (see Fig. \ref{fig:appcom}($a$) and ($b$)),
or, the Hubbard $U$ (see Fig. \ref{fig:appcom}(c) and (d)).

\section{The behavior of average sign in the Mott region}
\label{app:sign}

\begin{figure}[htb]
\begin{minipage}{4.2cm}
\includegraphics[width=4.9cm]{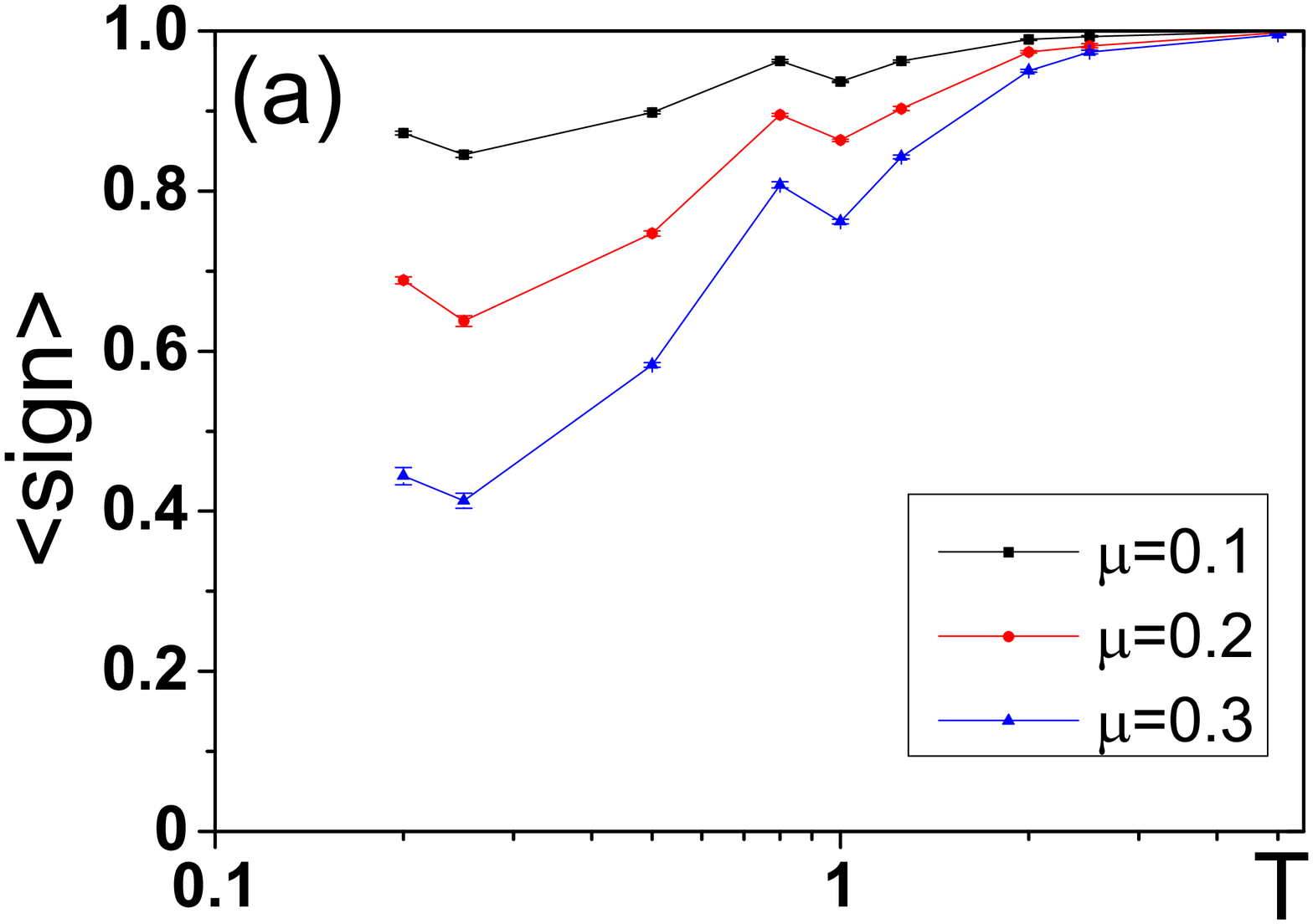}
\end{minipage}
\begin{minipage}{4.2cm}
\includegraphics[width=4.9cm]{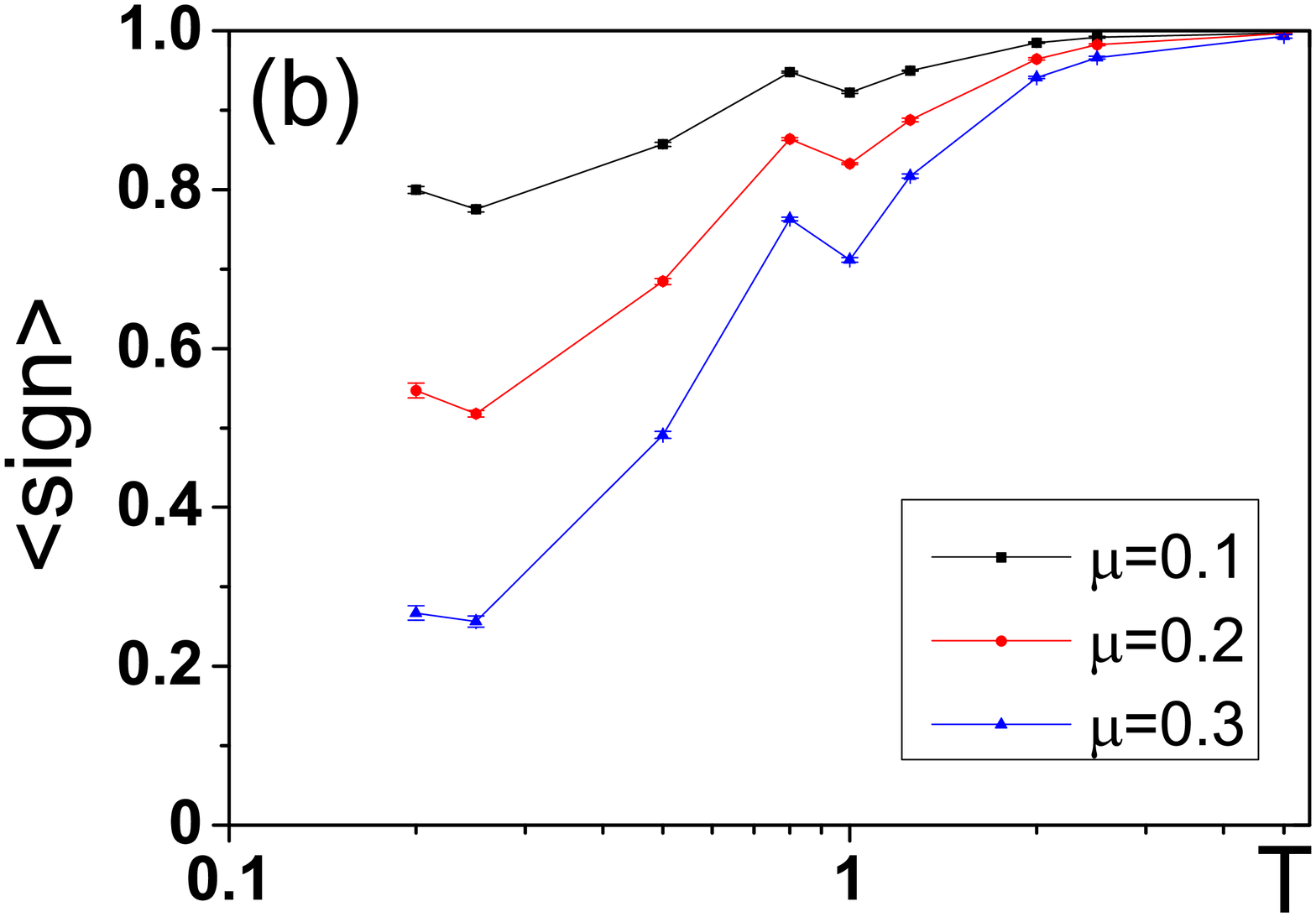}
\end{minipage}
\caption{ The average sign as a function of $T$ for various chemical potentials
at ($a$) $U=12$ and ($b$) $U=14$ in the SU($6$) Hubbard model on a honeycomb
lattice with $L=9$.
Error bars are smaller than the data points.
}
\label{fig:sign}
\end{figure}

The chemical potential is set to zero (at half-filling) in our DQMC simulations,
which ensures the sign problem is absent.
 We also test the average sign in the Mott region of the SU($6$) Hubbard model when the chemical potential $\mu$
 deviates from zero (away from half-filling). In this case, an extra term $H_{\mu}=-\mu \sum_{i} n_i$ is added to the original Hamiltonian, Eq. (\ref{eq:hubbard}).
As shown in Fig. \ref{fig:sign}, the average signs deviate quickly from unity
as the temperature decreases and the sign problem becomes severe when temperatures
are lower than $T\sim t$, a temperature scale set by the numerical instability test in Appendix \ref{app:com}.

\section{The AF structure factors and nearest-neighbor
spin-spin correlations }
\label{app:AF}
\begin{figure}[htb]
\begin{minipage}{4.2cm}
\includegraphics[width=4.9cm]{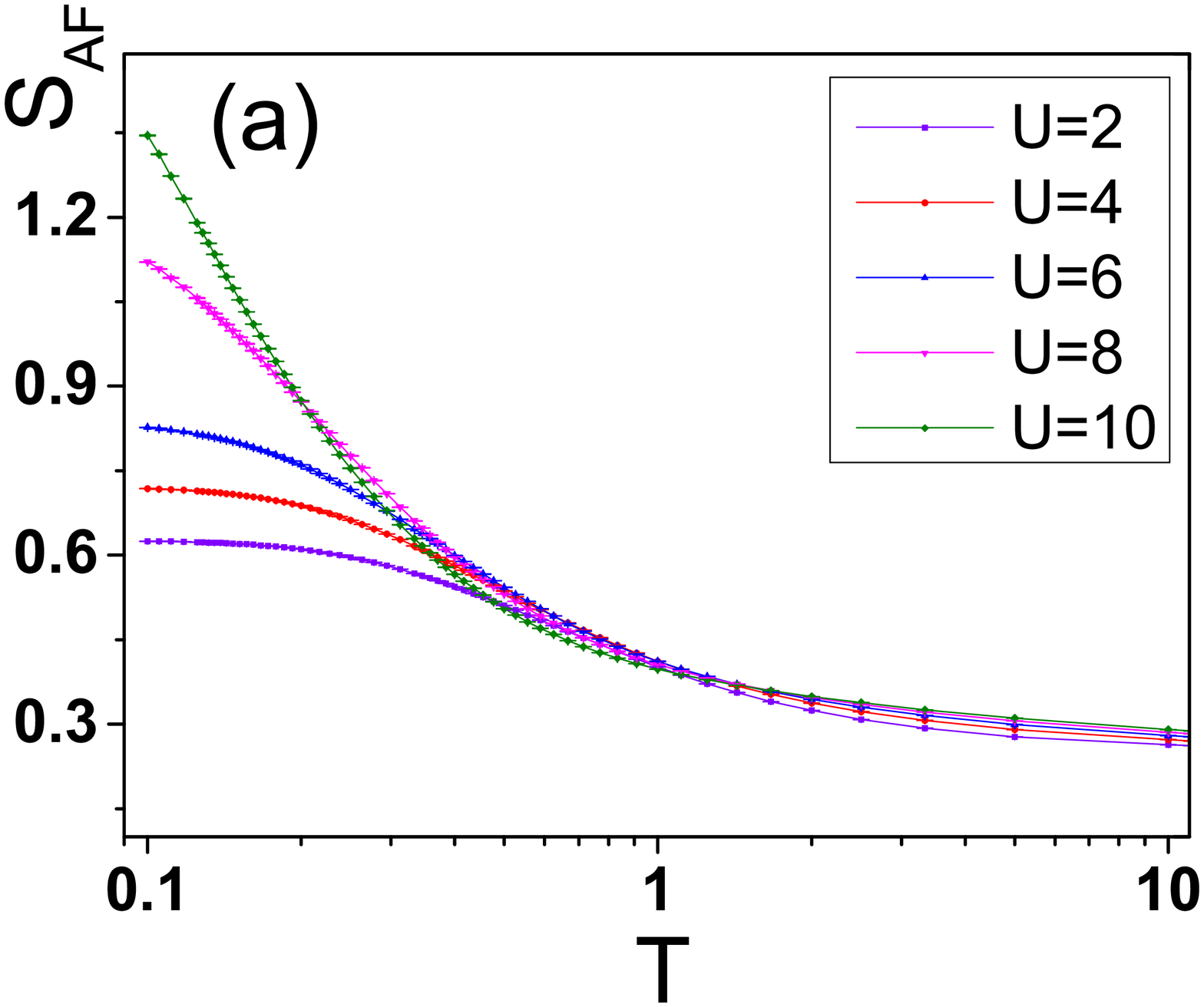}
\end{minipage}
\begin{minipage}{4.2cm}
\includegraphics[width=4.9cm]{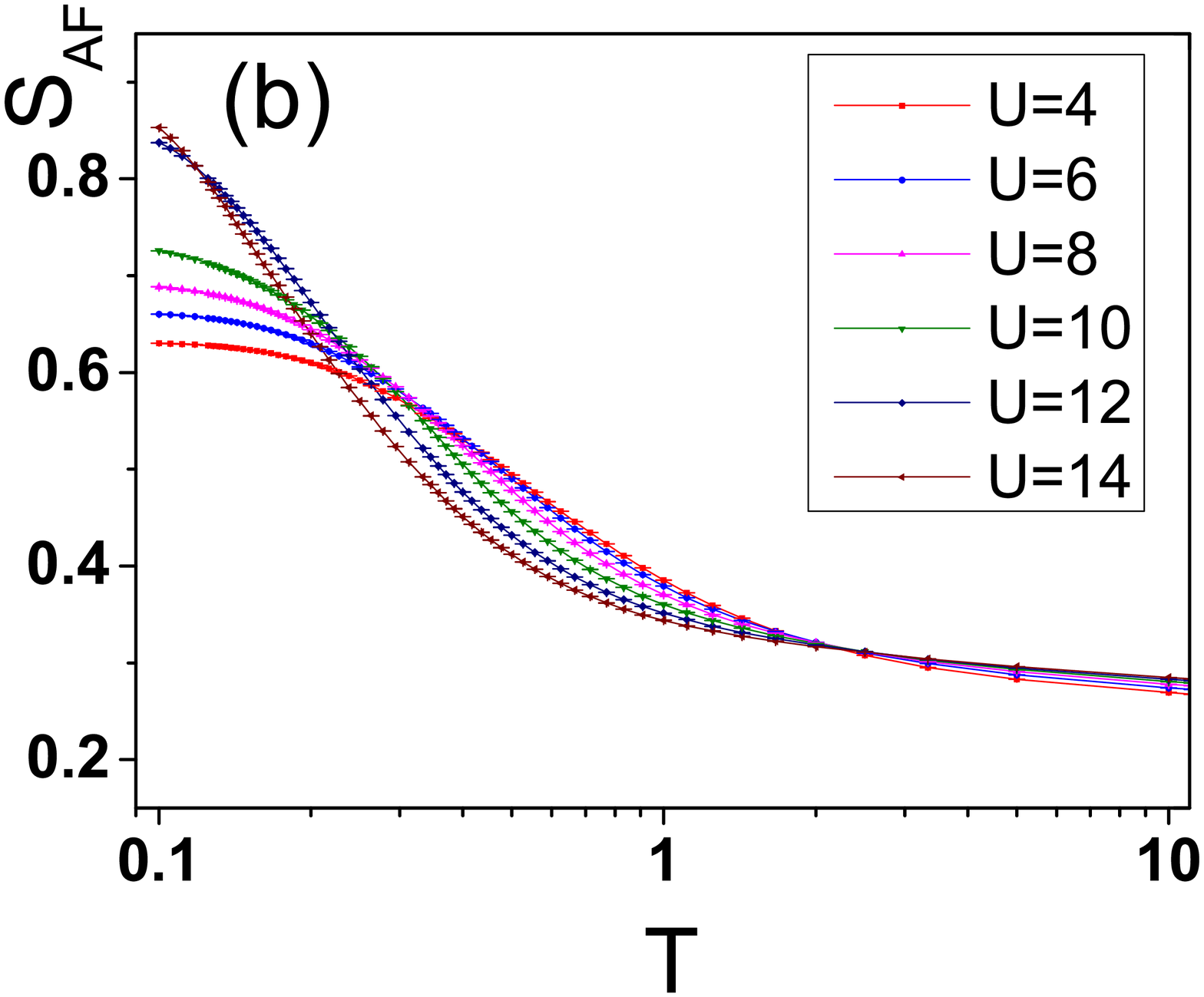}
\end{minipage}
\caption{The AF structure factors $S_{AF}$ versus temperature $T$ at various values of $U$
 in the half-filled
 ($a$) SU($4$) and ($b$) SU($6$) Hubbard models.
 The lattice size is $L=9$.
 Error bars are smaller than the data points.
}
\label{fig:TAF}
\end{figure}

\begin{figure}[htb]
\includegraphics[width=0.8\linewidth]{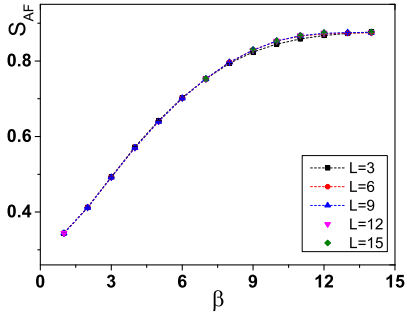}
\caption{The AF structure factor $S_{AF}$ of the SU($6$) Hubbard
  model with $U=14$ is plotted as a function
  of $\beta$ for different lattice sizes $L$.
  The dashed lines serve as a guide to the eye, and error bars are smaller
  than the data points.
}
\label{fig:Sorder}
\end{figure}

\begin{figure}[htb]
\begin{minipage}{4.2cm}
\includegraphics[width=4.9cm]{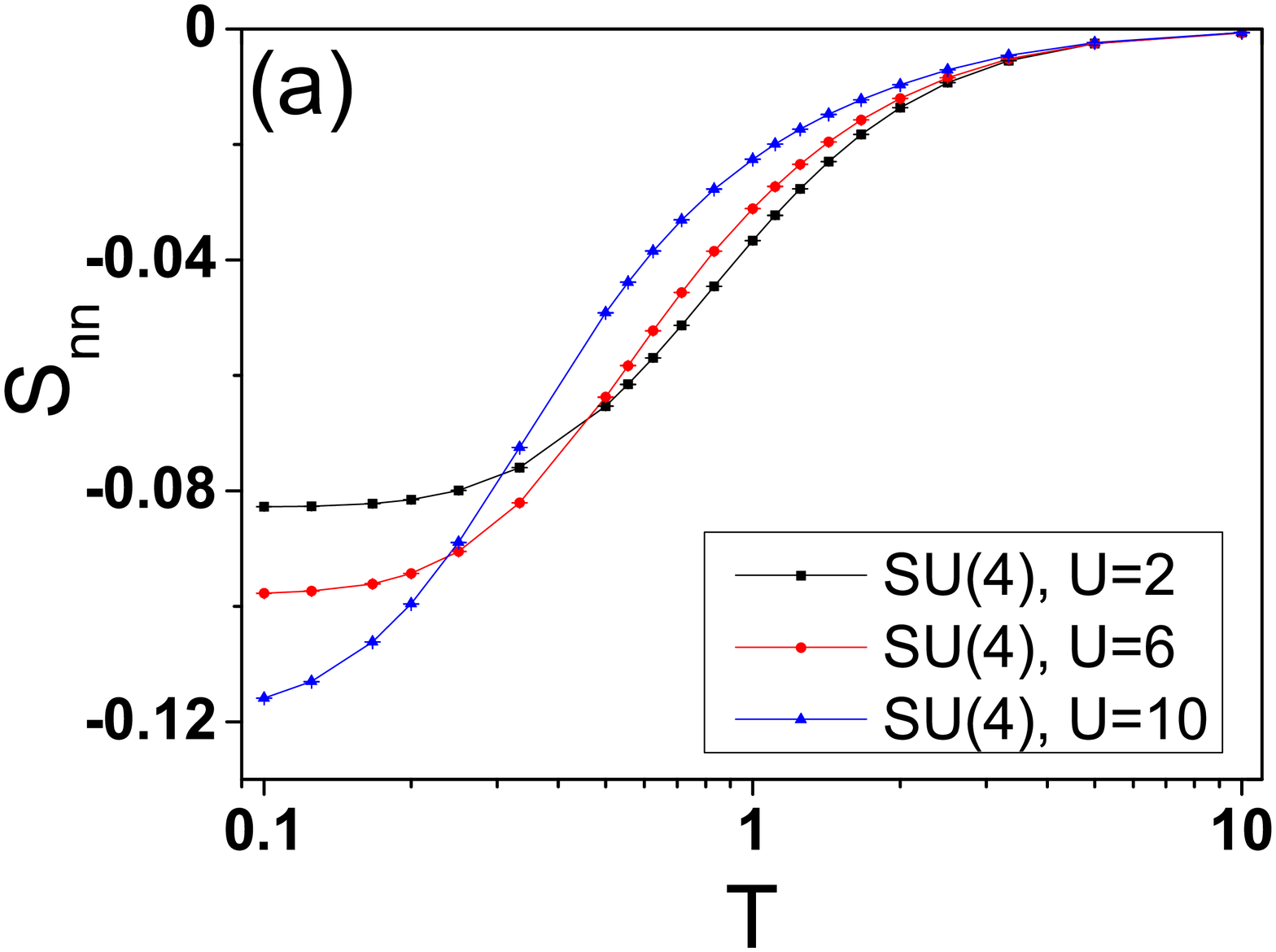}
\end{minipage}
\begin{minipage}{4.2cm}
\includegraphics[width=4.9cm]{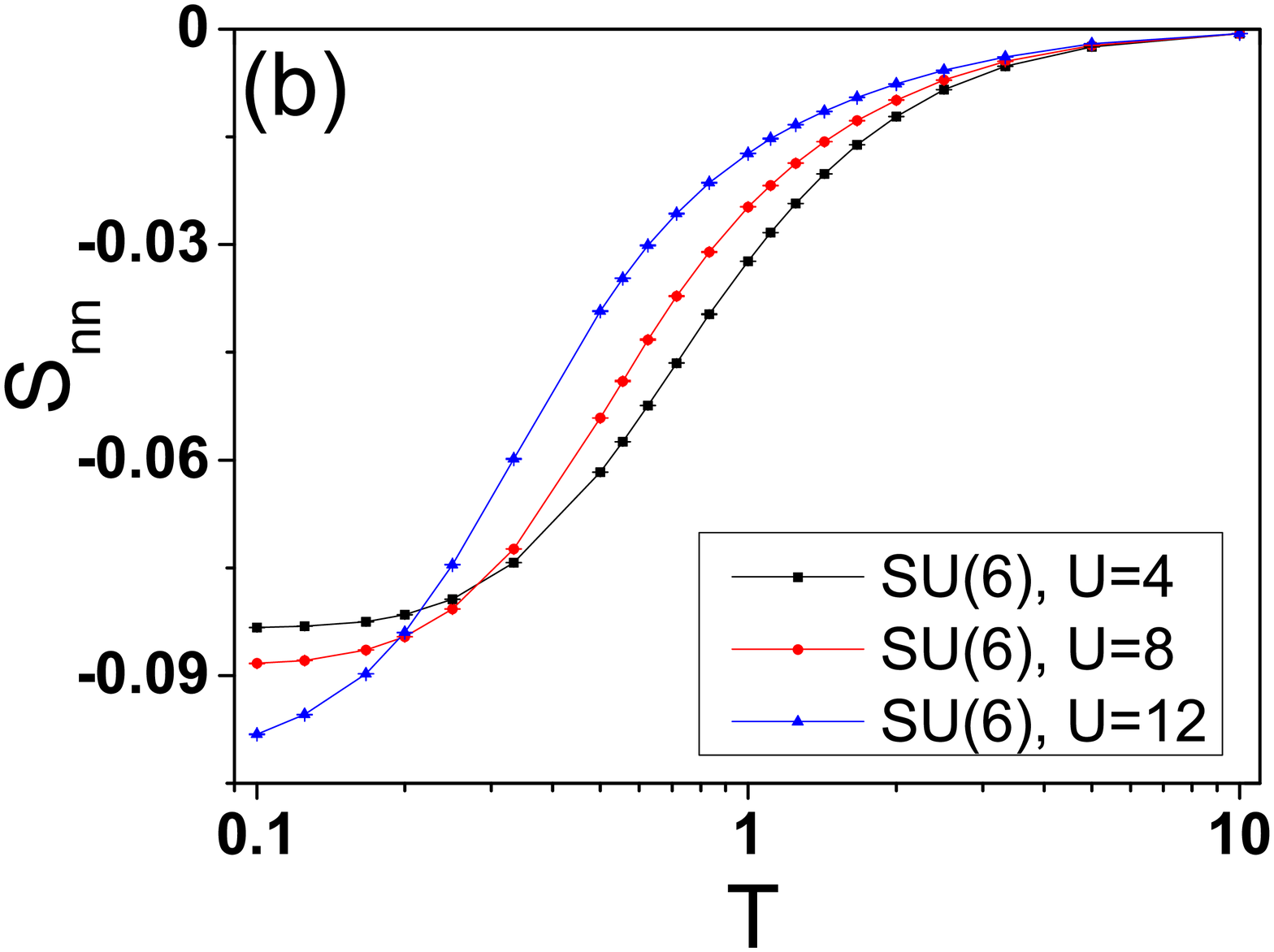}
\end{minipage}
\caption{The nearest-neighbor spin-spin correlation $S_{nn}$ versus temperature $T$ at various values of $U$
in the half-filled
($a$) SU($4$) and ($b$) SU($6$) Hubbard models.
The lattice size is $L=9$.
Error bars are smaller than the data points.
}
\label{fig:NNspin}
\end{figure}

The AF structure factors are defined as
\bea
S_{AF} = \frac{1}{2L^2}\sum_{i,j} (-1)^{i+j} S_{spin}(i,j).
\eea
As shown in Fig. \ref{fig:TAF}, we simulate the AF structure factor
$S_{AF}$ of the SU(4) and SU(6) Hubbard models on a $2 \times 9 \times 9$
honeycomb lattice.
With decreasing temperatures, the $S_{AF}$'s of the SU($2N$) ($N=2,3$) Dirac
fermions increase slowly and saturate eventually when $2\le U/t\le 6$,
while they increase rapidly when $8\le U/t\le10$.

In Fig. \ref{fig:Sorder}, the $\beta$ dependence of
the SU(6) AF structure factor $S_{AF}$
with $U/t=14$ are shown for different lattice sizes from $L=3$ to $L=15$.
$S_{AF}$ increases monotonically with inverse temperature $\beta$.
But $S_{AF}$ is size-independent even at low temperatures $T/t\sim 1/10$,
which indicates that the AF correlation length is smaller than
lattice size $L=3$.
This is another evidence that the long-range AF order is absent in
the half-filled SU($6$) Hubbard model on a honeycomb lattice.

The nearest-neighbor spin-spin correlations are defined as
\bea
S_{nn} = \frac{1}{zL^2} \sum_{i\in A, \vec{e}_{j}} S_{spin}(i,i+\vec{e}_{j}),
\eea
where $z$ is the coordination number.
In Fig. \ref{fig:NNspin}, we present the nearest-neighbor spin-spin
correlations $S_{nn}$ in the half-filled SU(4) and SU(6) Hubbard models.
At high temperatures $T/t\sim 10$, $|S_{nn}|$ is independent of the
Hubbard $U$, which shows spin incoherence.
In contrast, at low temperatures $T/t\sim 0.1$, increasing $U$
enhances the nearest-neighbor AF correlations, and thus
$|S_{nn}|$ increases.


%

\end{document}